\documentclass[fleqn,usenatbib]{mnras}

\usepackage{newtxtext,newtxmath}

\usepackage[T1]{fontenc}

\DeclareRobustCommand{\VAN}[3]{#2}
\let\VANthebibliography\thebibliography
\def\thebibliography{\DeclareRobustCommand{\VAN}[3]{##3}\VANthebibliography}

\usepackage{graphicx}	
\usepackage{amsmath}	
\newcommand{\Su}{\color{black}}
\newcommand{\Zhang}{\color{black}}

\title[Short title, max. 45 characters]{Mutual Occurrence Ratio of Planets. I. New Clues to Reveal Origins of Hot- and Warm-Jupiter from the RV Sample.}

\author[Su et al.]{
Xiang-Ning Su,$^{1,2,3,4}$
Hui Zhang,$^{2,3,4}$\thanks{E-mail: zhangh@shao.ac.cn}
Ji-Lin Zhou$^{3,4}$
\\
$^{1}$ School of Physics and Optoelectronic Engineering
, Hainan University, Haikou 570228, China\\
$^{2}$Shanghai Astronomical Observatory, Chinese Academy of Sciences, Shanghai 200030, China\\
$^{3}$School of Astronomy and Space Science, Nanjing University, Nanjing 210023, People’s Republic of China\\
$^{4}$Key Laboratory of Modern Astronomy and Astrophysics, Ministry of Education, Nanjing 210023, People’s Republic of China
}

\date{Accepted XXX. Received YYY; in original form ZZZ}

\pubyear{2022}

\begin{document}
\label{firstpage}
\pagerange{\pageref{firstpage}--\pageref{lastpage}}
\maketitle

\begin{abstract}
Many studies have analyzed planetary occurrence rates and their dependence on the host's properties to provide clues to planet formation, but few have focused on the mutual occurrence ratio of different kinds of planets. Such relations reveal whether and how one type of planet evolves into another, e.g. from a cold Jupiter to a warm or even hot Jupiter, and demonstrate how stellar properties impact the evolution history of planetary systems. 
We propose a new classification of giant planets, i.e. cold Jupiter(CJ), warm Jupiter(WJ), and hot Jupiter(HJ), according to their position relative to the snow line in the system. 
Then, we derive their occurrence rates(${\eta}_{\rm HJ}$, ${\eta}_{\rm WJ}$, ${\eta}_{\rm CJ}$) with the detection completeness of RV(Radial Velocity) surveys(HARPS$\&$ CORALIE) considered. 
Finally, we analyze the correlation between the mutual occurrence ratios, i.e. ${\eta}{_{\rm CJ}} / {\eta}_{\rm WJ}$, ${\eta}{_{\rm CJ}} / {\eta}_{\rm HJ}$ or ${\eta}{_{\rm WJ}}/{\eta}_{\rm HJ}$, and various stellar properties, e.g. effective temperature $T_{\rm eff}$.
Our results show that the ${\eta}_{\rm HJ}$, ${\eta}_{\rm WJ}$ and ${\eta}_{\rm CJ}$ are increasing with the increasing $T_{\rm eff}$ when $T_{\rm eff}\in (4600,6600] K$. 
Furthermore, the mutual occurrence ratio between CJ and WJ, i.e. ${\eta}{_{\rm CJ}} /{\eta}_{\rm WJ}$, shows a decreasing trend with the increasing $T_{\rm eff}$. But, both ${\eta}{_{\rm CJ}}/{\eta}_{\rm HJ}$ and ${\eta}{_{\rm WJ}}/{\eta}_{\rm HJ}$ are increasing when the $T_{\rm eff}$ increases.
Further consistency tests reveal that the formation processes of WJ and HJ may be dominated by orbital change mechanisms rather than the in-situ model. However, unlike WJ, which favors gentle disk migration, HJ favors a more violent mechanism that requires further investigation.
\end{abstract}

\begin{keywords}
 Exoplanets -- Planetary Systems
\end{keywords}



\section{Introduction}
The origins of hot and warm Jupiters (which are conventionally defined by their orbital period ranges, less than 10 days and from 10 to 200 days, respectively) are essential components to build up the general framework of the formation and evolution theory of giant planets. 
On one hand, hundreds of hot Jupiters have been discovered thanks to their short periods and large signals in RV (radial velocity) and transit observations. 
And this relatively large sample allows us to look at their formation process from a statistical perspective.
Although some people believe the hot Jupiter may form in-situ \citep{2016ApJ...829..114B, 2016ApJ...817L..17B, 2018ApJ...866L...2B}, most believe that these short-period giant planets should experience some kind of orbital migration scenarios: the disk migration, i.e. giant planets formed far away and migrated inward across the gaseous/dusty proto-stellar disk\citep{1980ApJ...241..425G, 1986ApJ...309..846L, 1996Natur.380..606L}, and the high-eccentricity migration, i.e. long-period giant planets may be excited to a high-eccentricity orbit by mechanisms like Lidov-Kozai effect\citep{1962AJ.....67..591K, 1962P&SS....9..719L} and further delivered to the close-in orbit by the tidal dissipation  \citep{1996Sci...274..954R, 1996Natur.384..619W, 2008ApJ...686..621F,  2008ApJ...686..580C}.
On the other hand, due to the longer orbital period, warm Jupiters are more difficult to be detected both by RV and transit methods, resulting in a relatively smaller sample size. 
And unlike hot Jupiters, warm Jupiters seem to be less difficult to form in-situ. 
Their origins and formation processes are still under intense debate. 
For example, is the rarity of warm Jupiters an observation bias or a result of low occurrence rate? 
Whether they have experienced some kind of orbital migration, or have they just formed in-situ? 
What's the relation between hot Jupiters and warm Jupiters?
According to the disk migration model, cold Jupiters may migrate inward to become warm Jupiters and even hot Jupiters. 
But recent studies challenged this according to the mass distribution of giant planets \citep{2020ApJ...904..134H}. 
While the violent high-eccentricity migration hypothesis has difficulties explaining the formation of warm Jupiters beyond 1 AU\citep{2016AJ....152..174A}. And according to the high companion fraction of warm Jupiters\citep{2016ApJ...825...98H, 2012PNAS..109.7982S}, their orbital evolution history should be gentle. 
Statistics on planetary occurrence rates are believed to be keys to answer such questions.
The measure of planetary occurrence rate requires a large and homogeneous planet sample. 
Thanks to the discoveries of the Kepler mission, many statistics studies have investigated the correlations between planetary occurrence rate and planetary dynamics properties, e.g., the overall planetary occurrence rate has an increasing trend with decreasing planetary radius and increasing orbital period \citep{2012ApJS..201...15H}. 
And others have focused on the correlation between the planetary occurrence rate and the property of the host star. \citet{2015ApJ...798..112M} and \citet{2020AJ....159..164Y} found that the occurrence rate of small planets (1–4 $R_{\oplus}$) is successively higher toward later spectral types at all orbital periods. For the dependence on the stellar metallicity, some people demonstrate that there is no discernible difference in the occurrence rate of medium-radius(2-4$R_{\oplus}$) and medium-period(10-100days) planets orbiting stars with different metallicity \citep{2018AJ....156..221N,2018AJ....155...89P}, while other studies have shown that the occurrence rate of Kepler planets slightly increases as the metallicity of host stars increases \citep{2020arXiv200308431K}. 

However, these statistics studies are based on the Kepler sample which is dominated by super-Earths and mini-Neptunes ($R_p\in[1,4] R_{\oplus}$) with moderate orbit periods ($p\leq300$days). 
Limited by the observation baseline, only K dwarfs harbor giant planets outside the snow line (corresponding to the equilibrium temperature of planets is 170K\citep{1981PThPS..70...35H}) in the Kepler sample. 
To investigate the formation processes of giant planets, especially the warm and cold giant planets, we need more long-period complimentary samples.
Radial velocity (RV) surveys have found many exoplanets with mass greater than that of Neptune  beyond 1AU\citep{2008PASP..120..531C,2010Sci...330..653H, 2010ApJ...709..396B,2011arXiv1109.2497M, 2014ApJ...785..126K,2015ApJ...800..138N,2016ApJ...819...28W, 2018ApJ...860..109G, 2020MNRAS.492..377W,2021ApJS..255....8R,2021ApJS..255...14F}. 
Based on RV-detected planets, some studies have also found an increasing occurrence rate of planets with decreasing planetary mass(3$M_{\oplus}$ to 1000$M_{\oplus}$) \citep{2010Sci...330..653H}, and a rising occurrence rate with orbital semi-major axis out to 1 AU  \citep{2013ApJ...778...53D,2016A&A...587A..64S,2020MNRAS.492..377W}, which are consistent with results based on Kepler planets. 
But other RV statistics studies suggest that the occurrence rate of giant planets appears to fall off after $~100$ day\citep{2019ApJ...874...81F} or 3 AU\citep{2021ApJS..255...14F}. 
And for the dependence on stellar properties, some studies based on RV planets have found that stars with higher metallicity and mass tend to host more giant planets\citep{2005ApJ...622.1102F,2010PASP..122..701J,2013A&A...551A..36N,2013A&A...551A.112M,2021ApJS..255...14F}. 

All results mentioned above are based on the present values of the planetary orbit, whose original distribution and dependency may have been reshaped substantially during the dynamics evolution history. 
The most straightforward way to solve this problem is taking into account the age of planet systems, i.e. age of their host stars. For example, statistical studies on the planetary occurrence rates versus stellar could demonstrate the evolution traces of planetary properties directly \citep{2020arXiv200514671B}.
However, due to the difficulties in obtaining accurate stellar ages of a large sample of planet-hosting stars, the valid sample size is still quite limited for reliable statistics studies\citep{2021AJ....161..189L}. 
In contrast to the direct way, studies on the ratio of occurrence rates of different planetary populations may offer us some indirect clues about their formation and evolution rates. 
For example, let us consider a large sample of planet-bearing proto-stellar disks whose initial properties are coupled strongly to the host star's properties, e.g. the disk mass is proportional to the stellar mass, effective temperature, and metallicity\citep{2016ApJ...831..125P,2021AAS...23731702M}, etc. 
Within each disk, we assume most giant planets form outside the snow line\citep{1996Icar..124...62P}(CJ) and some of them may migrate inward to become warm Jupiter (WJ) and even hot Jupiter(HJ). 
Then, the formation rate of WJ and HJ (evolved from CJ or WJ) when the disk has been completely depleted, should be governed by the migration efficiency within the disk, which is supposed to be highly correlated to the stellar properties, e.g. the stellar mass, effective temperature, and metallicity, etc. 
That means the ratio of occurrence rates of CJ, WJ, and HJ, should show a dependence on the stellar properties. 
If giant planets do not migrate at all, i.e. they formed just in-situ, the ratio of occurrence rates would just represent the relative probability of where a giant planet may emerge, e.g. inside or beyond the snow line. 
And such ratios may also show a dependence on the stellar properties, but are very different from the results from migrating model. 
Therefore, the stellar property-dependent ratio of occurrence rates (also known as mutual occurrence ratio) may be used to distinguish different formation and evolution mechanisms of different populations of giant planets, i.e. forming through disk migration or in-situ. 
In this work (the first paper of a series), we present our statistics studies on the dependency between the mutual occurrence ratios of three giant planet populations and their stellar properties, based on giant planet samples from RV surveys only.

This paper is arranged as follows: in the section \ref{sec:method} we describe the method of sample selection, our definitions of different giant planet populations in RV samples, and the occurrence rates of different populations of giant planets according to the detection completeness of corresponding surveys.
Major results are presented in section \ref{sec:results}. 
In section \ref{sec:discussion} and section \ref{sec:summary}, we discuss and conclude our results.

\section{Sample and Method} \label{sec:method}
\subsection{RV Samples and The Definitions of HJ, WJ, and CJ} \label{sec:boundary}
\begin{figure*}
	\centering
	\includegraphics[width =1 \textwidth]{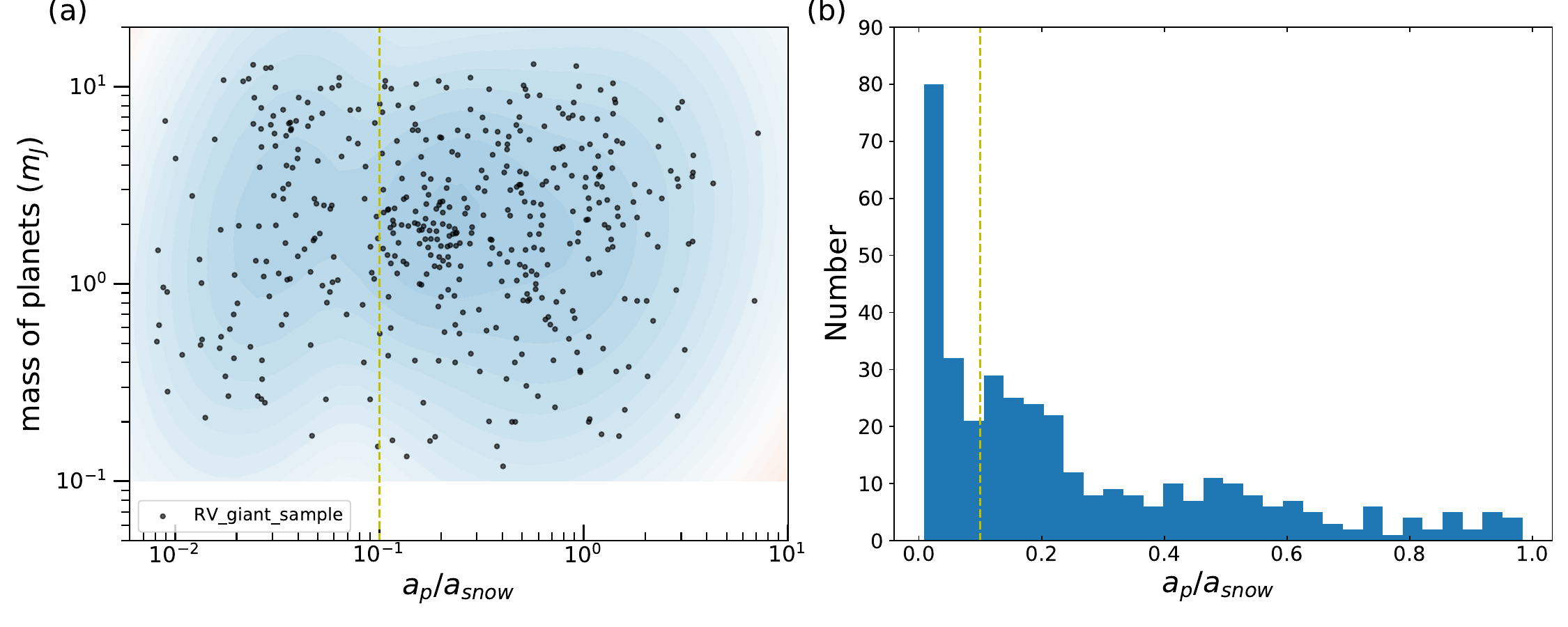}
	
	\caption{  The boundary between hot and warm Jupiters. (a) Distribution of  normalized semi-major axis(${a_{\rm p}}/a_{\rm snow}$) .vs. planetary mass ($M_{\rm J}$) of giant planets detected by RV surveys. According to the Gaussian Mixture Model(GMM), all the giant planets inside the snow line, i.e. ${a_{\rm p}}/a_{\rm snow}\leq1.0$, could be divided into two clusters by a boundary around ${a_{\rm p}}/a_{\rm snow}\sim0.08-1.0$. (b) Histogram of normalized semi-major axes of all the giant planets inside the snow line.  There is a valley around ${a_{\rm p}}/a_{\rm snow}\sim 0.1$.  This valley also existed in Kepler-detected giant planets candidates, More details are in the appendix \ref{sec:kepler}.  \label{fig:RV_b}}
\end{figure*}

To avoid systemic differences in planetary and stellar properties imposed by various fitting models adopted by different surveys and to derive reliable survey completeness, we focus on giant planets detected by the HARPS and CORALIE survey\citep{2011arXiv1109.2497M,2013A&A...551A.112M}. 
We start with a total of 1797 K, G, and F-type stars, of which host 131 planets. 
Then, all these stars are cross-matched with the Gaia DR2\citep{2018A&A...616A...1G} to get precise stellar properties, e.g. the effective temperature $T_{eff}$, stellar radius $R_{\star}$ and etc. 
Only stars with $T_{eff}$ between 4600K and 6600K are selected for further analysis. This is to avoid systematic errors caused by the known limitation of synthetic stellar atmospheric models \citep{2017AJ....154..107P}. The final RV star sample for deriving the planetary occurrence rates and the associated stellar property dependencies contains 1407 main sequence stars. 
For the RV planet sample, two more criteria are applied:
\begin{enumerate}
      \item select only confirmed giant planets within 0.1 and 13 Jupiter masses: $0.1M_{\rm J}<M_{\rm p}<13M_{\rm J}$.
      \item select giant planets with an orbital period shorter than 15000 days: $T_{\rm p} \leq 15000$ days, where the survey completeness obtained by the HARPS and CORALIE survey is valid \citep{2011arXiv1109.2497M}.
    \end{enumerate}
After using those filters, we get a homogeneous sample of 124 RV-detected giant planets. A new classification of these giant planets is as follows:
usually, we call a giant planet whose orbit period is less than a few days, e. g. $<10$ days, hot Jupiter. 
However, such a definition does not have too much physics meaning.
Planets with the same orbit period around different types of host stars may be significantly different in their physics and dynamics properties.
To reveal the stellar dependency of different planetary populations, we adopt a new definition according to the normalized orbit semi-major axis($a_{\rm p}/a_{\rm snow}$) which relies on both planetary and stellar properties at the same time.
The normalization factor is the position of the snow line $a_{\rm snow}$ where the equilibrium temperature $T_{\rm p}=170K$ in the planetary system. 
The equilibrium temperature of the planet is given by \cite{1981PThPS..70...35H}:
\begin{equation}\label{1}
T_{\rm p}=280(\frac{a_{\rm p}}{1AU})^{-1/2}(\frac{L_{\star}}{L_{\odot}})^{1/4} K
\end{equation}
where $L_{\star}$ is the luminosity of the host star, 
\begin{equation}\label{2}
L_{\star}=4\pi{R_{\star}}^{2}\sigma T_{\rm eff}^{4}
\end{equation}
where $R_{\star}$ is stellar radius, $\sigma$ is Stefan-Boltzmann constant, and $T_{\rm eff}$ is effective temperature of star. 
 As giant planets are believed to be more likely form beyond the snow line, it is naturally to classify those giant planets with semi-major axis larger than the snow line $a_{\rm snow}$ (${a_{\rm p}}/a_{\rm snow}>1.0$) as cold Jupiters. However, determining the boundary between warm and hot Jupiters is much more complicated. The diagram of normalized semi-major axis(${a_{\rm p}}/a_{\rm snow}$) versus planetary mass($M_{\rm J}$) of all known RV detected giant planets is shown in Figure \ref{fig:RV_b}(a). Based on the Bayesian Information Criterion(BIC), we derive the optimal fitting model of the Gaussian Mixture Model(GMM) to find the boundaries between potential clusters. This method has been well used to analyze the significance of gaps in the distribution of planetary masses\citep{2019ApJ...880L...1A}. As shown in Figure \ref{fig:RV_b}(a), we find all the RV detected giant planets could be marginally divided into two clusters along the x-axis with a boundary around ${a_{\rm p}}/a_{\rm snow}\sim 0.08-0.1$. We check the histogram of the normalized semi-major axis distribution, which also reveals the presence of a valley close to ${a_{\rm p}}/a_{\rm snow}\sim 0.1$ (Figure \ref{fig:RV_b}(b)). We also analyse the normalized semi-major axis distribution 
of Kepler-detected giants planets candidate, there is also have a valley around ${a_{\rm p}}/a_{\rm snow}\sim 0.1$(see in appendix \ref{fig:kepler}).Thus, we define those giant planets with normalized semi-major axis not larger than 0.1, i.e. ${a_{\rm p}}/a_{\rm snow}\leq0.1$, as hot Jupiter (HJ). And the warm Jupiter (WJ) are those with a normalized semi-major axis between 0.1 and 1, $0.1<{a_{\rm p}}/a_{\rm snow}\leq1.0$. 

\begin{figure*}
	\centering
	\includegraphics[width =1.0 \textwidth]{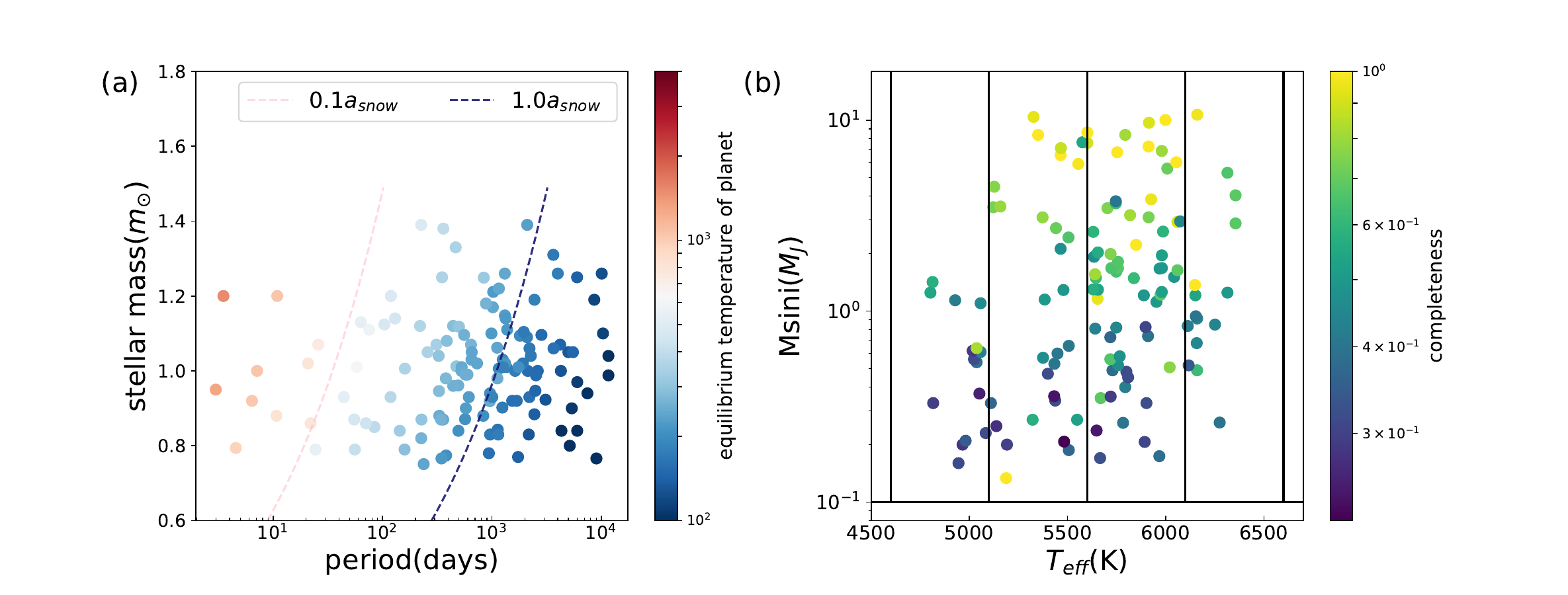}
	
	\caption{Overview of our samples. (a): the distribution of orbital period and stellar mass(p-$M_{*}$) of all samples, the colors denote the equilibrium temperature of each giant planet (from blue to red corresponds to 100 K to
	4000K, where pink dotted line and dark blue dotted line mark the position of $0.1a_{\rm snow}$  and $1.0 a_{\rm snow}$ of different host stars; (b): the
	distribution of planetary mass and effective temperature of host star( $m_{\rm p}$-$T_{\rm eff}$) of our giant planets samples, detection completeness shown as color contours from 0.0 to
1.0 in steps of 0.1. }
\label{fig:f1}
\end{figure*}

The final HJ, WJ, and CJ sample each contains 13, 77, and 34 giant planets respectively. 
Figure \ref{fig:f1}(a) shows the distribution of the orbital period of our samples versus the mass of their host star. The color of the solid dots indicates the equilibrium temperature of the planet in logarithmic scale. It is clear that the planetary equilibrium temperature strongly depends on both the planet orbit period and stellar mass. Based on this new classification, we derive the planetary occurrence rate of each of three planetary populations and investigate the dependence on the stellar properties.

\subsection{Occurrence Rates} \label{sec:Occurrence}
Many correlations between planetary occurrence rate and various stellar properties, most of which are of stellar mass and metallicity, have been reported before. 
Here we select the stellar effective temperature, $T_{\rm eff}$, as a major representative stellar property. 
There are two reasons to select the effective temperature: First, it is an essential parameter to derive the stellar mass, the spectral type, and furthermore, a bridge to the properties of the proto-stellar disk. 
Second, the measurement of effective temperature is easier to achieve higher accuracy than that of stellar mass and other properties. 
As mentioned above, we have narrowed $T_{\rm eff}$ of our RV samples down between $4600K$ to $6600K$. 
Then we divide all samples into four bins with a uniform bin size of 500K \citep{2012ApJS..201...15H}. 
The occurrence rate of a specific type of giant planet in each effective temperature bin is derived as following:
\begin{enumerate}
\item In each effective temperature bin, we count the number of stars, ${N}_{\star}$. The number of stars within each of the four bins is listed in table \ref{tab:RV_label}.

\item For each RV-detected giant planet, we derive the survey completeness, ${p}_{\rm j}$, based on its mass $M_{\rm p}\sin{i}$ and orbit period $P_{\rm p}$. The $M_{\rm p}-P_{\rm p}$ dependent survey completeness of HARPS and CORALIE survey is provided by \citet{2011arXiv1109.2497M} (in Figure 6).  For each star in the HARPS and CORALIE survey, the probability of detecting a planet with a specific period ${p}$ and minimum mass $M_{\rm p}\sin{i}$ was calculated. To determine the overall survey completeness, these probabilities were averaged across all stars in the survey and offered by \citet{2011arXiv1109.2497M}. We linearly interpolate their completeness curves into a uniform grid with $M_{\rm p}\sin{i}$ between 0.1 and 13 $M_{\rm J}$ and a period between 1 and 15000 days. Then, the survey completeness of a sample is the average value of the grid where it falls. 
The color map in Figure \ref{fig:f1}(b)) shows the survey completeness to each planet with the brighter color denoting higher completeness. 
 Because most of giant planets in our samples have eccentricities below 0.5, the survey completeness estimated by \citet{2011arXiv1109.2497M} does not consider the influences caused by the orbital eccentricity. To address this issue, we carried out a series of inject-and-recover tests\citep{2021AJ....162..272S} for RV signals with random orbit eccentricities, and we find that it does not affect the RV detectability significantly (which is similar to previous results, e.g. \citet{2002A&A...392..671E}, \citet{2010MNRAS.401.1029C},\citet{2016ApJ...819...28W}). 
 However, we discuss the effects of all non-zero orbital eccentricities in the Appendix \ref{sec:eccentricity}, taking the periastron and apastron distances into consideration. We first taken eccentricity into account in the classification of
 planets, as well as its effect on the occurrence rate of giant planet.

\item Calculate the average number of giant planets per star in each effective temperature bin by considering the survey completeness:
\begin{equation}\label{3}
{\eta}= \frac{1}{{N}_{\star}}\sum_{j=1}^{{n}_{\rm p}}\frac{1}{{p}_{\rm j}},
\end{equation}
where ${n}_{\rm p}$ is the total detected number of planets in each effective temperature bin of the RV survey(shown in table \ref{tab:RV_label}). And we assume the uncertainty in the occurrence rate follows the Poisson distribution.

\item Finally, when we got the occurrence rate of HJ, WJ and CJ (which are denoted as ${\eta}_{\rm HJ}$, ${\eta}_{\rm WJ}$, and ${\eta}_{\rm CJ}$, respectively), we further derive the mutual occurrence ratios:${\eta}{_{\rm CJ}}/{\eta}_{\rm WJ}$, ${\eta}{_{\rm CJ}}/{\eta}_{\rm HJ}$ and ${\eta}{_{\rm WJ}}/{\eta}_{\rm HJ}$, in each effective temperature bin. The uncertainty on the mutual occurrence rate is calculated from the error transfer formula\citep{w9100796}.
\end{enumerate}

{\Su
\begin{table*}
    \centering
    \caption{Number of stars and giant planet in RV -detected samples}
    \begin{tabular}{p{5cm} p{4cm} p{5cm}}
    \hline
    Effective Temperature(K) & Number of stars(${N}_{\star}$) & Number of giant planets(${n}_{\rm p}$) \\
    \hline
    (4600, 5100] & 314 & 15\\
    (5100, 5600] & 454 & 30\\
    (5600, 6100] & 510 & 64\\
    (6100, 7100] & 129 & 15 \\
    \hline
    \end{tabular}
    \label{tab:RV_label}
\end{table*}
}
\section{Results} \label{sec:results}
\begin{figure*}
	\centering
	\includegraphics[width =1.0 \textwidth]{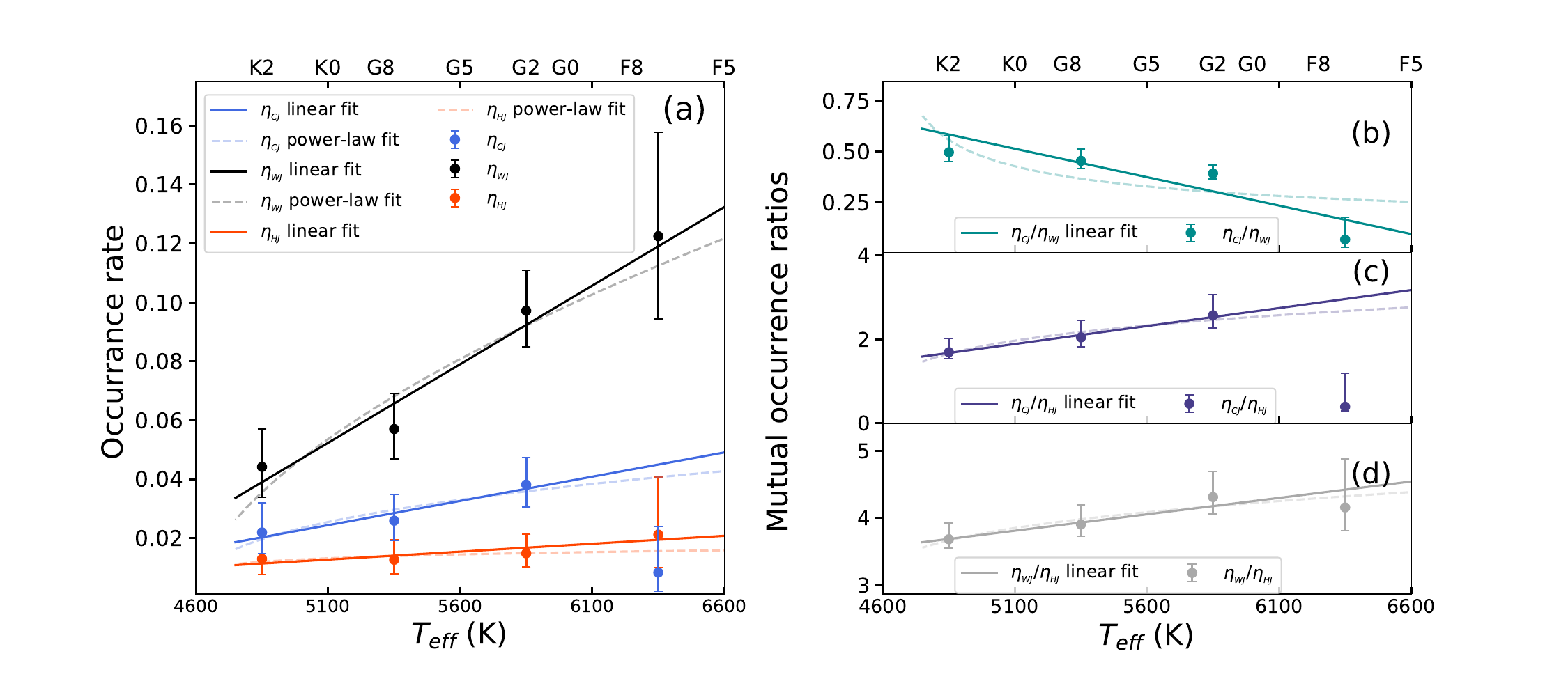}
	
	\caption{Occurrence rate and mutual occurrence rate of different giant planet populations(HJ, WJ, CJ) in RV survey. (a): The occurrence rate of giant planets in four $T_{\rm eff} $ bins ($T_{\rm eff}\in[4600K,6600K]$),HJ(red), WJ (black) and CJ (blue) are depicted separately, the error bars indicate the $68.3\%$ confidence intervals. The best-fit linear occurrence model for HJ, WJ, and CJ is shown as red, black, and blue solid lines. And 
dashed curves are power-law fit. (b),(c),(d): the mutual occurrence ratio of giant planets, green dots indicate ${\eta}{_{\rm CJ}} / {\eta}_{\rm WJ}$, ${\eta}{_{\rm CJ}} / {\eta}_{\rm HJ}$ are marked by purple dots, gray dots indicate ${\eta}{_{\rm WJ}}/{\eta}_{\rm HJ}$, solid lines of the same color represent the best linear fit of those mutual occurrence ratios, and the power-law fit is represented by dashed curves.}
\label{fig:RV_all}
\end{figure*}

\subsection{Occurrence rates Versus Stellar $T_{\rm eff}$}

We investigate the correlation between planetary occurrence rate and the effective temperature of the host stars. 
Within each $T_{\rm eff} $ bin, we derive the planetary occurrence rates: ${\eta}_{\rm HJ}$, ${\eta}_{\rm WJ}$, and ${\eta}_{\rm CJ}$. 
And their dependencies on the stellar effective temperature have been shown in panel (a) of Figure \ref{fig:RV_all}. 
In the range of $4600-6600K$, all the three occurrence rates, ${\eta}_{\rm HJ}$, ${\eta}_{\rm WJ}$ and ${\eta}_{\rm CJ}$, show an increasing trend when the stellar effective temperature $T_{\rm eff}$ is increasing. 
The increasing slope is different for different planet populations. 
The solid lines in Figure \ref{fig:RV_all} show the best fits to this relation by utilizing the optimizing algorithm in the SciPy package: 
\begin{equation}\label{fit_1}
f(T_{\rm eff})={{\eta}_0}+k(\frac{T_{\rm eff}-T_{0}}{1000}),
\end{equation}
where $T_{0}=4600K$, the best-fit coefficients are, for HJ, ${{\eta}_{\rm 0,H}}=0.011 \pm 0.002$, $k_{\rm H}=0.006 \pm 0.002$; for WJ, ${{\eta}_{\rm 0,W}}=0.028 \pm 0.009$, $k_{\rm W}=0.069 \pm 0.01$ and for CJ, ${{\eta}_{\rm 0,C}}=0.017 \pm 0.004$, $k_{\rm C}=0.023 \pm 0.005$.

Note that ${\eta}_{\rm CJ}$ drops downward sharply and has a large uncertainty in the rightmost bin where $T_{\rm eff} > 6100K$. This is caused by the lack of long-period giant planets around early-type stars which requires tens of years' observation baseline. And the other possible reason, which may be more essential, is that early-type stars (e.g. F-type) whose effective temperature is higher are relatively more difficult to get high precision radial velocity measurements. As a result, only those quiet early-type stars are selected in an RV planet survey, leading to a smaller sample size of higher $T_{\rm eff}$ stellar targets. And this additional selection bias has not been considered in the detection efficiency correction part. In fact, we have marked the stellar spectrum types corresponding to a stellar effective temperature in panel (a) of Figure \ref{fig:RV_all}. The number of giant planets in the $6100-6600K$ bin (including early G and late F stars) is relatively rare, especially for the CJ, where only 1 sample is covered. So the uncertainties of ${\eta}_{\rm CJ}$ is relatively larger and this value of ${\eta}_{\rm CJ}$ in this bin needs further validation. We perform tests on alternative boundaries for giant planets and observed that the occurrence rate of Jupiter is not sensitive to the choice of boundary as the effective temperature increases(Appendix \ref{sec:ac}).

First, if we sum the occurrence rates of all three kinds of giant planets in each stellar effective temperature bin, we find the occurrence rate of giant planets increases with stellar effective temperature. For the main sequence star (according to Fig.1 in \citet{2011arXiv1109.2497M}, our star samples are in the main sequence phase), the effective temperature is proportional to the stellar mass (The correspondence between the mass and the effective temperature of our star sample is shown in the Appendix\ref{sec:a1}). This means a giant planet favors a massive star, which is consistent with previous results, e.g. \citet{2010PASP..122..701J}.  

Second, from the best-fit results of the $T_{\rm eff}-{\eta}$ relation, we find that ${\eta}_{\rm WJ}$ is always larger than ${\eta}_{\rm HJ}$ and ${\eta}_{\rm CJ}$. It is not a surprise to see ${\eta}_{\rm WJ}> {\eta}_{\rm HJ}$, since the orbit range of WJ is much wider than that of HJ. However, the result of ${\eta}_{\rm WJ}>{\eta}_{\rm CJ}$ for most KGF stars (where $T_{\rm eff}\in[4400,6600]K$) is difficult to be explained by the standard in-situ formation theory since it is easier to build a giant planet outside the snow line where the surface density of solid material jumps much higher than that inside the snow line. In contrast, this is a natural result if most of CJ would finally become WJ or even HJ by the inward orbit change caused by the disk migration or high-e migration mechanisms. And furthermore, it seems that ${\eta}_{\rm WJ}$ grows faster towards higher $T_{\rm eff}$ than ${\eta}_{\rm CJ}$ does. This implies that the rates of such orbit change may be more effective when the star is hotter and heavier. This subtle difference in the ratio of ${\eta}_{\rm CJ}/{\eta}_{\rm WJ}$ provides a way to further distinguish the detailed migration processes, disk migration, or high-e.

\subsection{mutual occurrence ratios of HJ, WJ and CJ}
To investigate the efficiency of different transitions between HJ, WJ and CJ, we derived the ratio of occurrence rates (or mutual occurrence ratio), e.g. ${\eta}{_{\rm CJ}}/{\eta}_{\rm WJ}$, ${\eta}{_{\rm CJ}}/{\eta}_{\rm HJ}$ and ${\eta}{_{\rm WJ}}/{\eta}_{\rm HJ}$, and its dependency on the stellar effective temperature. Instead of absolute occurrence rate, such relative ratios also mitigate those statistical effects caused by the characteristics of the mother sample. The uncertainty in the ratio of occurrence rates is derived from the uncertainties of each occurrence rate following the error transfer formula\citep{w9100796}. 

As shown in the panel (b) of Figure \ref{fig:RV_all}, there is a monotonous decreasing trend of the mutual occurrence ratio of CJ and WJ with the increasing stellar effective temperature, i.e. $d({\eta}{_{\rm CJ}}/{\eta}_{\rm WJ})/dT_{\rm eff} < 0$, the slope of the best fit curve is $ -0.32\pm 0.01$. This implies that the inward orbital migration should be more and more efficient when the host star becomes hotter. If the inward orbital migration of CJ is dominated by the disk, then this would be a natural result. Because, for the main sequence star, a hotter effective temperature always means a higher stellar mass, and usually also means it had a more massive proto-stellar disk. To efficiently deliver a giant planet from outside the snow line to inside, the cradle disk where the planet is embedded should be massive enough to carry away a large amount of angular momentum from the planet. And the more massive is the disk, the more efficient the migration is. 

As most HJ are unlikely to form in-situ, they must be delivered from some places further away through some mechanisms, e.g. the gentle migration through disk or the violent high-eccentricity scattering. In other words, WJ and CJ are potential predecessors of HJ. So, we also investigate the other mutual occurrence ratios, ${\eta}{_{\rm CJ}} / {\eta}_{\rm HJ}$ and ${\eta}{_{\rm WJ}} / {\eta}_{\rm HJ}$, which would indicate the transition efficiency of HJ from either CJ or WJ. In contrast to ${\eta}{_{\rm CJ}} / {\eta}_{\rm WJ}$, both ${\eta}{_{\rm CJ}} / {\eta}_{\rm HJ}$ and ${\eta}{_{\rm WJ}} / {\eta}_{\rm HJ}$ show an increasing trend with increasing stellar $T_{\rm eff}$. The trend is almost monotonous, except for the drop at the bin of $6100-6600K$, where the excessive uncertainty has been discussed in the previous section. Furthermore, compared to the ${\eta}{_{\rm WJ}} / {\eta}_{\rm HJ}$, the growth trend of ${\eta}{_{\rm CJ}} / {\eta}_{\rm HJ}$ is more gradual. Those significant difference in the $T_{\rm eff}$-dependency of the mutual occurrence ratios indicates that the formation and evolution mechanisms of HJ and WJ may be quite different. Some further analyses are as follows.

\section{Discussions}\label{sec:discussion}
\subsection{Sample Selection} 
With the development of observation technology, the detection accuracy of radial velocity is constantly improved. Our samples were published in 2011\citep{2011arXiv1109.2497M}. In 2013, it was applied to analyze the relationship between the occurrence rate of planets and metallicity of stars\citep{2013A&A...551A.112M}. We get star samples from table 1 and table 2 of article\citep{2013A&A...551A.112M}. The typical precision ($\sim 5m/s$) of CORALIE radial velocity measurements is sufficient for giant planet observations, and the precision of HARPS is $\sim 1m/s$ (on a 7.5 magnitude star). In addition, our sample has two advantages: on the one hand, the host stars are mostly non-active main-sequence stars, and on the other hand, most giant planets have eccentricities of less than 0.5. However, there are new RV surveys database published in last year: The California Legacy Survey (CLS; \citep{2021ApJS..255....8R,2021ApJS..255...14F}, including 719 FGKM stars, and 178 exoplanets). Therefore, we also conducted a similar filtrate on CLS data. But, some problems are as follows: (1) Some stars are not in the main sequence phase. (2) There are no cold Jupiters in $T_{\rm eff}\in (6100,6600] K$ bin according to our criteria, even though the survey completeness is applicable to planets in 0.03-30 AU. 

 We use the Gaia DR2 data to obtain host stars' parameters. While the parameters derived from the latest data show improvements, our findings indicate that the utilized data currently falls within the margin of error of the most recent data (Gaia DR3). Moreover, the discrepancies with data inferred through spectroscopic analysis are even smaller. Therefore, we opted to continue utilizing the previous dataset. A detailed comparison of the results can be found in Appendix B.
\subsection{Implication to the formation of HJ and WJ} 

The formation of giant planets located inside the snow line, i.e. HJ and WJ, is usually considered to be an outside-in process (for most HJ and a part of WJ at least). 
Giant planets form outside the snow line, and then they migrate inward to become WJ or even HJ. 
There are two major mechanisms to explain the inward delivery of giant planets, i.e. the migration through a disk which is composed of gas, dust, or/and planetesimals, and the migration due to planet-planet scattering which is also called the high-eccentricity (high-e) migration. 

One of the major goals of most recent statistical studies about giant planets is to find out the dominant formation mechanism and the major evolution track. In this work, we study the mutual occurrence ratio of three kinds of giant planets, i.e. ${\eta}{_{\rm CJ}} / {\eta}_{\rm WJ}$, ${\eta}{_{\rm WJ}} / {\eta}_{\rm HJ}$ and ${\eta}{_{\rm CJ}} / {\eta}_{\rm HJ}$. And we believe these outer-to-inner ratios and their dependence on the stellar effective temperature will reveal useful clues on the efficiency of transition from CJ to WJ (or HJ) and from WJ to HJ, when the host star is of different stellar types.

Our results show that the transition efficiency trends of WJ ($d(\eta_{\rm CJ}/ \eta_{\rm WJ})/dT_{\rm eff}$) and HJ ($d(\eta_{\rm CJ}/\eta_{\rm HJ})/dT_{\rm eff}$, $d(\eta_{\rm WJ}/\eta_{\rm HJ})/dT_{\rm eff}$) are significantly different. This difference implies the formation mechanisms of WJ and HJ should be quite different, since the result associated with WJ could be naturally explained by the gentle disk-migration mechanism, while the result associated with HJ is contrary to the prediction from the same mechanism. However, before we could jump to the conclusion, there are still two consistency issues that need to be addressed and clarified, i.e. influences of the stellar metallicity and the prediction of in-situ formation mechanism.

\begin{figure*}
	\centering
	\includegraphics[width =1.0 \textwidth]{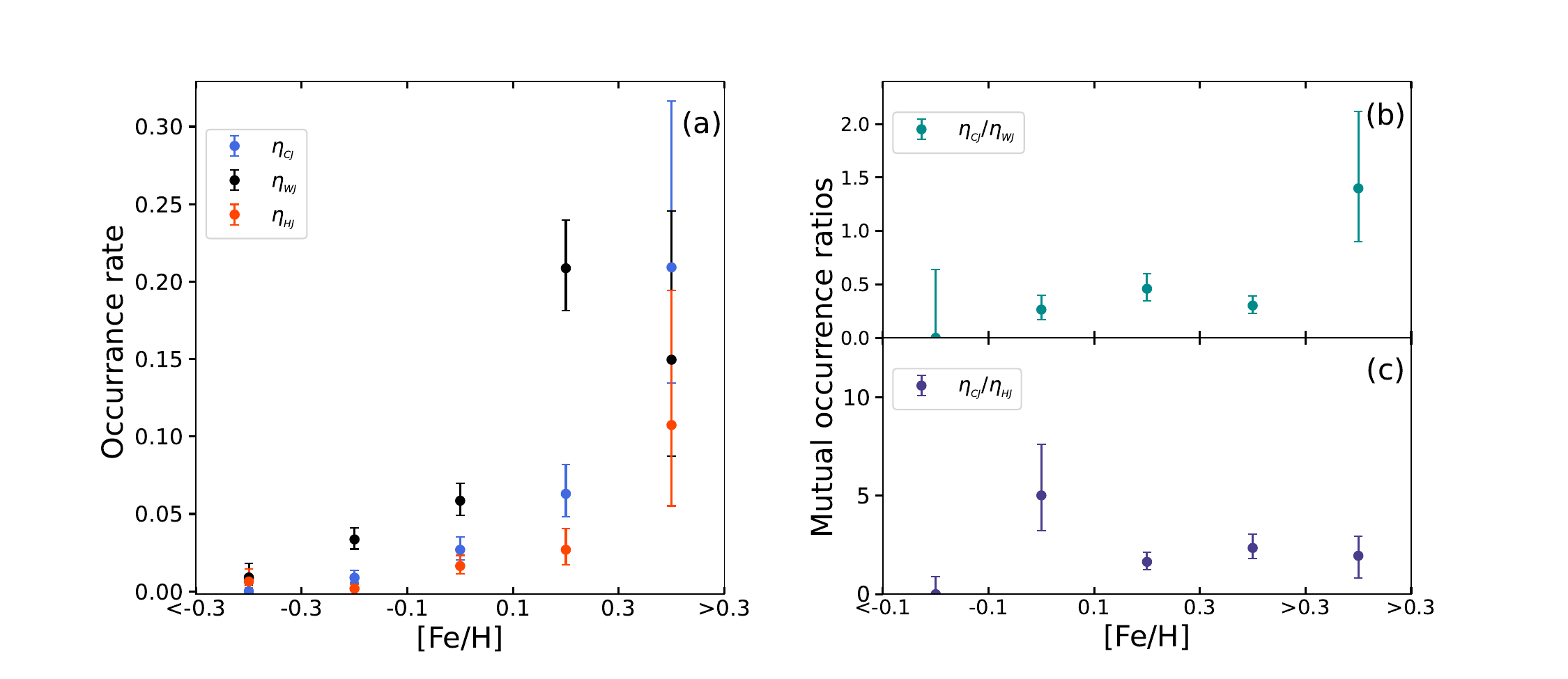}
	
	\caption{The occurrence-rate- stellar metallicity dependence of CJ, WJ, and HJ, and the mutual occurrence ratios of CJ, WJ, and HJ. (a): the occurrence rate of giant planets as a function of stellar metallicity, similar to Figure \ref{fig:RV_all}, the red(black, blue) dots show the occurrence rate of HJ(WJ, CJ). (b): the mutual occurrence ratios ${\eta}{_{\rm CJ}}/{\eta}_{\rm WJ}$(shown in green) as a function with stellar metallicity.(c): the mutual occurrence ratio ${\eta}{_{\rm CJ}}/{\eta}_{\rm HJ}$ (shown in purple) as a function with stellar metallicity.  
	\label{fig:RV_m}}
\end{figure*}

The correlation between the occurrence rate of giant planets and the metallicity of the host star has drawn much attention and been treated as a critical clue to reveal the roles of stellar properties in the planetary formation processes \citep{2010PASP..122..701J,2013A&A...551A.112M,2015ApJ...798..112M,2020arXiv200308431K,2021ApJS..255...14F}. It is believed that giant planets, especially the long-period ones, e.g. WJ and CJ, are more likely to form around metal-rich stars \citep{2010PASP..122..701J,2013A&A...551A..36N}. To perform a sanity check and verify the consistency of our model, we also explore the stellar-metallicity dependence of HJ, WJ, and CJ. The definition, sample selection, and method to derive the planetary occurrence rate are all the same as that of our $T_{\rm eff}$ related studies. The only difference is we replace the stellar effective temperature with the stellar metallicity [Fe/H]. As shown in panel (a) of Figure \ref{fig:RV_m}, all three occurrence rates increase with an increasing metallicity of the host star. This result is consistent with previous studies\citep{2010PASP..122..701J,2013A&A...551A..36N,2021ApJS..255...14F}. Furthermore, the metallicity dependencies of ${\eta}{_{\rm CJ}} / {\eta}_{\rm WJ}$ and ${\eta}{_{\rm CJ}} / {\eta}_{\rm HJ}$ are also different: ${\eta}{_{\rm CJ}} / {\eta}_{\rm WJ}$ shows an increasing trend with the increasing metallicity (in panel (b) of Figure \ref{fig:RV_m}), while ${\eta}{_{\rm CJ}} / {\eta}_{\rm HJ}$ follows a decreasing trend with increasing metallicity in the general view (panel (c) of Figure \ref{fig:RV_m}).

According to the core-accretion model, it is believed that most giant planets would form slightly outside the snow line where the solid density jumps high due to the condensation of water vapor. Before a planetary core has accreted enough solid mass to trigger further gas accretion effectively, it usually suffers faster inward orbital decay due to the tidal torques exerted on it by the gaseous disk. It is usually called the type-I migration which always causes a significant orbit change within $10^4-10^5$ year for a core mass around several $M_\oplus$. Many planetary cores may be delivered to a warm or even hot place from their cold birthplace outside the snow line by this mechanism. 
As soon as the planet core grows above a critical mass which depends on the disk opacity and is usually around $\sim 10M_\oplus$, it steps into a quasi-static accretion stage. 
At this stage, it takes several million years to reach a total mass (of the solid core and the gaseous shell) of above $\sim 20M_\oplus$, and then steps into the runaway gas accretion scenario\citep{1996Icar..124...62P}. This would build up a massive giant planet in a very short time scale, $\sim 10^3-10^4$ years. 
After the planet has grown massive enough, it will open a deep gap in the disk along its orbit region and the orbital migration rate will drop down significantly to match the viscous evolution time-scale of the disk, i.e. the type-II migration. 
While, the high metallicity of the proto-stellar disk may help to accelerate the growth of the solid planet core through the pebble accretion mechanism and shorten, even avoid the fast type-I migration era. 

If we assume the metallicity of the proto-stellar disk is directly proportional to the metallicity of the host star, giant planets may grow more quickly around a metal-rich star and stay close to the original place where they are born, e.g. slightly beyond the snow line. In other words, the transition efficiency from a CJ to a WJ will be lower around a star with higher metallicity. And the mutual occurrence ratio of CJ and WJ, ${\eta}{_{\rm CJ}} / {\eta}_{\rm WJ}$, should increase as the stellar metallicity increases, which is just consistent with our result (see panel (b) in Figure \ref{fig:RV_m}). While, in contrast, the decreasing trend of ${\eta}{_{\rm CJ}} / {\eta}_{\rm HJ}$ implies a transition process from CJ to HJ other than the disk-migration theory. This result is consistent with that of the $T_{\rm eff}$-related analyses.

All previous discussions have assumed that a giant planet origin beyond the snow line and may migrate inward as it is growing up and ends up as a warm or hot Jupiter. Then what would happen, if the cold, warm, or even hot giant planet forms in-situ? We perform some further analyses as follows.

\subsection{The hypothesis of In-situ formation}
\begin{figure*}
	\centering
	\includegraphics[width =0.7 \textwidth]{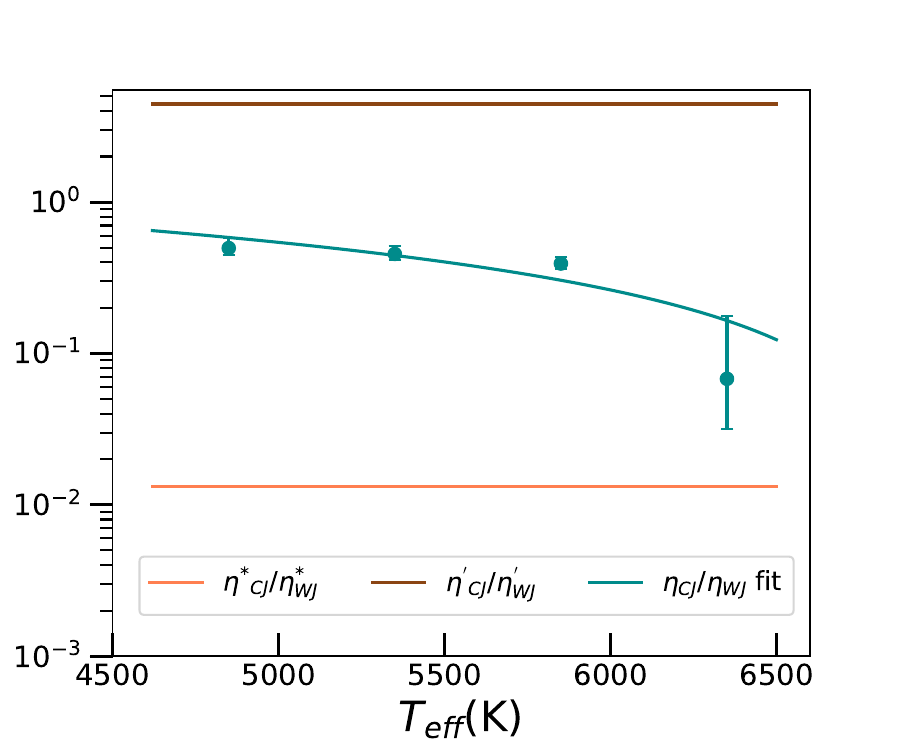}
	
	\caption{The comparison of the simulation and the observational result of mutual occurrence-rate-$T_{\rm eff}$ dependence of CJ and WJ. The green curve is the best fit of the result (Figure \ref{fig:RV_all}(b)). The brown curve and the orange curve show the simulation result of the ratio of occurrence rate based on the position-independent and position-dependent hypotheses, respectively.
	\label{fig:sim_RV}}
\end{figure*}

We start with the assumption that the formation efficiency of a giant planet is position-independent, which means the probability that a giant planet may show up is uniform within a planetary system. 
This is obviously not true but could be used as a reference. In this case, giant planets distribute uniformly in space and the odd of a giant planet becoming either a WJ or a CJ depends only on the position of the snow line. 
In other words, the ratio of formation rates ${{\eta}'}{_{\rm CJ}}/ {{\eta}'}_{\rm WJ}$ is the ratio of the distance from the snow line to the outer boundary of the system to the distance from the central host to the snow line. 
Since our definitions on HJ, WJ and CJ depend on the position of snow line, $a_{\rm p}\leq0.1 a_{\rm snow}$, $0.1a_{\rm snow} < a_{\rm p} \leq 1.0a_{\rm snow}$ and $a_{\rm p} > 1.0 a_{\rm snow}$, respectively, we set the outer boundary to $5 a_{\rm snow}$. 
Then, as the position of the snow line, $a_{\rm snow}$, moves outward with the increasing stellar effective temperature, the outer boundary also increases. 
As a result, the ratio ${{\eta}'}{_{\rm CJ}}/ {{\eta}'}_{\rm WJ}$ should be a constant and won't change with stellar type in this case (See the upper horizontal solid brown line in Figure \ref{fig:sim_RV}). 

In a more realistic case, we instead assume the efficiency to form a giant planet is position-dependent, which means the probability where a giant planet may show up depends on the surface density ------the higher the surface density is, the easier it is to form a giant planet. Therefore, the formation rates are proportional to the potential mass of planetary building blocks inside and outside the snow line, ${{\eta}^*}{_{\rm CJ}}$, ${{\eta}^*}_{\rm WJ}$. The formation rates are determined by the function of solid and gaseous mass density and the growth timescale, which are:
\begin{equation}\label{5}
{{\eta}^*}{_{\rm WJ}}=\mathop{ \int }\nolimits_{{0.1a_{\rm snow}}}^{{1.0a_{\rm snow}}} \frac{2\pi x (\Sigma_{\rm g}(x)+\Sigma_{\rm d}(x))}{t_{\rm acc}(x)} \text{d} x
\end{equation}

\begin{equation}\label{6}
{{\eta}^*}{_{\rm CJ}}=\mathop{ \int }\nolimits_{{1.0a_{\rm snow}}}^{{5.0a_{\rm snow}}} \frac{2\pi x (\Sigma_{\rm g}(x)+\Sigma_{\rm d}(x))}{t_{\rm acc}(x)} \text{d} x,
\end{equation}

where $a_{\rm snow}$ is the position of the snow line where the equilibrium temperature $T_{\rm p} = 170K$(equation (\ref{1})):
\begin{equation}\label{7}
a_{\rm snow}=
(\frac{280}{170})^2 (\frac{L_{\star}}{L_{\odot}})^{1/2} AU, 
\end{equation}
for each star of a given mass $M_{*}$, its luminosity is $L_{\star}=1.03{M_{*}}^{3.42}$, its radius is $R_{\star}=0.85{M_{*}}^{0.67}$\citep{1991Ap&SS.181..313D},according to equation (\ref{2}), the effective temperature of the star is $T_{\rm eff}\sim {M_{*}}^{0.52}$.And $t_{\rm acc}$ is the growth timescale of planetary embryos \citep{2004ApJ...616..567I}:
\begin{equation}\label{8}
t_{\rm acc}(x)\sim {\Sigma_{\rm d}}^{-1}{M_{*}}^{-1/2}x^{3/2}yr,
\end{equation}
$\Sigma_{\rm d}$ and $\Sigma_{\rm g}$ are the surface density of dust and gas, respectively, within the proto-stellar disk:

\begin{equation}\label{9}
\Sigma_{\rm d}(x)=f_{\rm d} f_{\rm snow} \times 10(\frac{x}{1AU})^{-3/2} g cm^{-2},
\end{equation}
\begin{equation}\label{10}
\Sigma_{\rm g}(x)=f_{\rm g} (2.4\times10^3)(\frac{x}{1AU})^{-3/2} g cm^{-2}.
\end{equation}

We adopt the stellar-mass dependent $f_{\rm d}=0.7{M_{*}}^{1.9}$, while fix $f_{\rm snow}=1$(inside the snow line) and $f_{\rm snow}=4.2$(outside the snow line). And we set $f_{\rm d}=f_{\rm g}$ to represent the solar abundance. So, the formation rates ${{\eta}^*}{_{\rm CJ}}$ and ${{\eta}^*}_{\rm WJ}$ are
\begin{eqnarray}\label{11}
{{\eta}^*}{_{\rm CJ}} \sim 1.59\times 10^{-4
}{T_{\rm eff}}^{0.048} 
\end{eqnarray}
\begin{eqnarray}\label{12}
{{\eta}^*}_{\rm WJ} \sim 1.20\times 10^{-2}{T_{\rm eff}}^{0.048} 
\end{eqnarray}

The ratios, ${{\eta}^*}{_{\rm CJ}} / {{\eta}^*}_{\rm WJ}$, around stars with different effective temperatures are shown in Figure\ref{fig:sim_RV} (the orange solid line). It turns out that this mutual occurrence ratio does not depend on the stellar effective temperature, $T_{\rm eff}$, which is determined by the stellar mass and has been canceled in the calculation. In fact, if giant planets form in-situ and do not travel across the snow line, the mutual occurrence ratio will not depend on any stellar property. 
And if the mutual occurrence ratio shows some dependencies upon stellar properties, it is a piece of strong evidence for large-scale mass exchange across the snow line, i.e. the orbit migration via various mechanisms.

\subsection{Consistency with previous works}
Although previous studies have not discussed the correlation between the \emph{mutual occurrence ratio} of different types of giant planets and stellar properties, many studies have investigated the correlation between the single occurrence rate of a certain type of giant planet and stellar parameters. To ensure consistency with previous studies, we also calculated the individual occurrence rate of three types of giant planet as a function of stellar effective temperature. And we found that there is an increasing trend in occurrence rate with increasing stellar effective temperature. Since the majority of our star sample is at the main sequence stage, the stellar effective temperature could be treated as a proxy for stellar type and mass with a lower measuring uncertainty (See Appendix \ref{sec:a1}). 
Our result is consistent with previous studies based on giant planet samples found by radial-velocity surveys \citep{2005ApJ...622.1102F,2010PASP..122..701J,2013A&A...551A..36N,2013A&A...551A.112M,2018ApJ...860..109G,2020MNRAS.491.5248W,2021ApJS..255....8R,2021ApJS..255...14F}, i.e. the occurrence of giant planets is positively correlated with both stellar mass (or stellar effective temperature) and metallicity. \citet{2010PASP..122..701J} and \citet{2020MNRAS.491.5248W} discovered a growth in the occurrence rate of giant planets in correlation with the incremental rise in stellar mass, commencing around 1 $M_{\odot}$and extending beyond it. \citet{2021ApJS..255...14F} from The California Legacy Survey(CLS; \citep{2021ApJS..255....8R,2021ApJS..255...14F}, including 719 FGKM stars, and 178 exoplanets) also confirmed this conclusion. Our result further extends the aforementioned conclusion to encompass stars with masses exceeding 0.7$M_{\odot}$. Based on the findings from the sample of evolved stars(the “retired A stars”) and an updated dataset of 1225 FGKM dwarfs, \citet{2018ApJ...860..109G} provide further evidence supporting the assertion that higher metallicity and mass tend to host more giant planets. Our result confirms those conclusions for giant planet around stars in the main sequence stage.

\section{Conclusions}\label{sec:summary}
In this paper, we try to explore the formation processes of different kinds of giant planets statistically. To classify different giant planets with more physics meanings related to their host star, we propose a new definition of hot Jupiter (HJ), warm Jupiter (WJ), and cold Jupiter (CJ), according to the planetary position relative to the snowline of the system, i.e. HJ: $a_{\rm p} \leq 0.1a_{\rm snow}$, WJ: $0.1a_{\rm snow}<a_{\rm p} \leq a_{\rm snow}$, and CJ: $a_{\rm p}>a_{\rm snow}$. Under this classification, we build a sample of RV-detected giant planets based on the HARPS $\&$ CORALIE surveys. Then, With this homogeneous sample, we investigate the individual occurrence rate of three populations of giant planets, i.e. ${\eta}_{\rm HJ}$, ${\eta}{_{\rm WJ}}$, and ${\eta}_{\rm CJ}$, the mutual occurrence ratios, i.e. ${\eta}{_{\rm CJ}} / {\eta}_{\rm WJ}$, ${\eta}{_{\rm CJ}} / {\eta}_{\rm HJ}$and ${\eta}{_{\rm WJ}} / {\eta}_{\rm HJ}$, and their dependency on the host's stellar properties, e.g. effective temperature and metallicity.

Firstly, we find that ${\eta}_{\rm HJ}$, ${\eta}{_{\rm WJ}}$, and ${\eta}_{\rm CJ}$ all show an increasing trend with the increasing stellar effective temperature, $T_{\rm eff}$. The slopes of their trend are significantly different:  ${\eta}{_{\rm WJ}}$ is always higher than ${\eta}{_{\rm CJ}}$ and ${\eta}{_{\rm HJ}}$, and it grows even faster toward higher $T_{\rm eff}$. This fact implies that the formation of WJ may be dominated by some kind of orbit migration process (see Section 3.1 and 4.2).

Secondly, we find that there are strong dependencies between the mutual occurrence ratios and the effective temperature of the host star. ${\eta}{_{\rm CJ}} / {\eta}_{\rm WJ}$ shows a declining trend with the increasing $T_{\rm eff}$, which can be explained by the orbit migration through the proto-stellar disk where higher $T_{\rm eff}$ leads to higher transition efficiency from CJ to WJ. While, the other two mutual occurrence ratios, ${\eta}{_{\rm CJ}} / {\eta}_{\rm HJ}$ and ${\eta}{_{\rm WJ}} / {\eta}_{\rm HJ}$ show a rising trend with increasing $T_{\rm eff}$ of the host star. In contrast, this implies that the formation of HJ should be very different than that of WJ, i.e. HJ is not simply the consequence of fast migrated WJ (see Section 3.2).

Thirdly, to check the consistency of our model, we also check the correlations between occurrence rates and the stellar effective temperature and our results are consistent with previous similar studies. And the mutual occurrence ratio, ${\eta}{_{\rm CJ}} / {\eta}_{\rm WJ}$ increases toward higher stellar metallicity. This is also consistent with the prediction made by the disk-migration theory and implies that the formation process of WJ is dominated by the gentle disk-migration process. Furthermore, the declined trend of ${\eta}{_{\rm CJ}} / {\eta}_{\rm HJ}$ toward higher metallicity also tells us that the formation of HJ is unlikely the same as that of WJ. Again, WJ and HJ should have experienced very different orbit change mechanisms (see Section 4.1). 

Finally, to ensure that the orbit change, which is either caused by the disk-migration or the high-e migration, has dominated the formation process of WJ and HJ, we test how the mutual occurrence ratio, ${{\eta}^*}{_{\rm CJ}} / {{\eta}^*}_{\rm WJ}$, would depend on the stellar effective temperature when both WJ and HJ form in-situ. Our calculations show that if there is no mass exchange across the snow line, i.e. large scale orbit migration of giant planets, then the mutual occurrence ratio, ${{\eta}^*}{_{\rm CJ}} / {{\eta}^*}_{\rm WJ}$, should be independent to the stellar properties, e.g. the stellar mass, effective temperature, etc. And the obvious correlation between such mutual occurrence ratios and the stellar properties is a piece of strong evidence of the orbit migration history of giant planets. 

In summary, our results show that the formation process of WJ and HJ is dominated by large-scale orbit migration instead of the in-situ mechanism. But WJ and HJ favor different orbit change mechanisms, i.e. most WJ should have experienced gentle disk migration, while most HJ should form through a more violent way like high-e migration. In this paper, we only present the results based on RV-detected giant planets. We will further present a similar result derived from the Kepler sample in our next paper.

\section*{Acknowledgements}

This work is supported by the support by the National Natural Science Foundation of China (NSFC; grant No. 12073010, 12303078, 11673011, 11933001, 12373043, 12063001). X.-N.S. also acknowledges the supported from Hainan Provincial Natural Science Foundation of China(424QN216, 122RC546, 124CXTD422). We thank Prof Rob Wittenmyer for helpful discussions.
{\Zhang We thank the reviewers for their constructive comments and suggestions}. 

\section*{Data Availability}

The data underlying this article will be shared on reasonable request to the corresponding author.



\bibliographystyle{mnras}
\bibliography{example} 

\begin{thebibliography}{}
\makeatletter
\relax
\def\mn@urlcharsother{\let\do\@makeother \do\$\do\&\do\#\do\^\do\_\do\%\do\~}
\def\mn@doi{\begingroup\mn@urlcharsother \@ifnextchar [ {\mn@doi@} {\mn@doi@[]}}
\def\mn@doi@[#1]#2{\def\@tempa{#1}\ifx\@tempa\@empty \href {http://dx.doi.org/#2} {doi:#2}\else \href {http://dx.doi.org/#2} {#1}\fi \endgroup}
\def\mn@eprint#1#2{\mn@eprint@#1:#2::\@nil}
\def\mn@eprint@arXiv#1{\href {http://arxiv.org/abs/#1} {{\tt arXiv:#1}}}
\def\mn@eprint@dblp#1{\href {http://dblp.uni-trier.de/rec/bibtex/#1.xml} {dblp:#1}}
\def\mn@eprint@#1:#2:#3:#4\@nil{\def\@tempa {#1}\def\@tempb {#2}\def\@tempc {#3}\ifx \@tempc \@empty \let \@tempc \@tempb \let \@tempb \@tempa \fi \ifx \@tempb \@empty \def\@tempb {arXiv}\fi \@ifundefined {mn@eprint@\@tempb}{\@tempb:\@tempc}{\expandafter \expandafter \csname mn@eprint@\@tempb\endcsname \expandafter{\@tempc}}}

\bibitem[\protect\citeauthoryear{{Antonini}, {Hamers}  \& {Lithwick}}{{Antonini} et~al.}{2016}]{2016AJ....152..174A}
{Antonini} F.,  {Hamers} A.~S.,   {Lithwick} Y.,  2016, \mn@doi [\aj] {10.3847/0004-6256/152/6/174}, \href {https://ui.adsabs.harvard.edu/abs/2016AJ....152..174A} {152, 174}

\bibitem[\protect\citeauthoryear{{Armstrong}, {Meru}, {Bayliss}, {Kennedy}  \& {Veras}}{{Armstrong} et~al.}{2019}]{2019ApJ...880L...1A}
{Armstrong} D.~J.,  {Meru} F.,  {Bayliss} D.,  {Kennedy} G.~M.,   {Veras} D.,  2019, \mn@doi [\apjl] {10.3847/2041-8213/ab2ba2}, \href {https://ui.adsabs.harvard.edu/abs/2019ApJ...880L...1A} {880, L1}

\bibitem[\protect\citeauthoryear{{Bailey} \& {Batygin}}{{Bailey} \& {Batygin}}{2018}]{2018ApJ...866L...2B}
{Bailey} E.,  {Batygin} K.,  2018, \mn@doi [\apjl] {10.3847/2041-8213/aade90}, \href {https://ui.adsabs.harvard.edu/abs/2018ApJ...866L...2B} {866, L2}

\bibitem[\protect\citeauthoryear{{Batygin}, {Bodenheimer}  \& {Laughlin}}{{Batygin} et~al.}{2016}]{2016ApJ...829..114B}
{Batygin} K.,  {Bodenheimer} P.~H.,   {Laughlin} G.~P.,  2016, \mn@doi [\apj] {10.3847/0004-637X/829/2/114}, \href {https://ui.adsabs.harvard.edu/abs/2016ApJ...829..114B} {829, 114}

\bibitem[\protect\citeauthoryear{{Berger}, {Huber}, {Gaidos}, {van Saders}  \& {Weiss}}{{Berger} et~al.}{2020}]{2020arXiv200514671B}
{Berger} T.~A.,  {Huber} D.,  {Gaidos} E.,  {van Saders} J.~L.,   {Weiss} L.~M.,  2020, arXiv e-prints, \href {https://ui.adsabs.harvard.edu/abs/2020arXiv200514671B} {p. arXiv:2005.14671}

\bibitem[\protect\citeauthoryear{{Boley}, {Granados Contreras}  \& {Gladman}}{{Boley} et~al.}{2016}]{2016ApJ...817L..17B}
{Boley} A.~C.,  {Granados Contreras} A.~P.,   {Gladman} B.,  2016, \mn@doi [\apjl] {10.3847/2041-8205/817/2/L17}, \href {https://ui.adsabs.harvard.edu/abs/2016ApJ...817L..17B} {817, L17}

\bibitem[\protect\citeauthoryear{{Bowler} et~al.,}{{Bowler} et~al.}{2010}]{2010ApJ...709..396B}
{Bowler} B.~P.,  et~al., 2010, \mn@doi [\apj] {10.1088/0004-637X/709/1/396}, \href {https://ui.adsabs.harvard.edu/abs/2010ApJ...709..396B} {709, 396}

\bibitem[\protect\citeauthoryear{{Chatterjee}, {Ford}, {Matsumura}  \& {Rasio}}{{Chatterjee} et~al.}{2008}]{2008ApJ...686..580C}
{Chatterjee} S.,  {Ford} E.~B.,  {Matsumura} S.,   {Rasio} F.~A.,  2008, \mn@doi [\apj] {10.1086/590227}, \href {https://ui.adsabs.harvard.edu/abs/2008ApJ...686..580C} {686, 580}

\bibitem[\protect\citeauthoryear{{Cumming} \& {Dragomir}}{{Cumming} \& {Dragomir}}{2010}]{2010MNRAS.401.1029C}
{Cumming} A.,  {Dragomir} D.,  2010, \mn@doi [\mnras] {10.1111/j.1365-2966.2009.15634.x}, \href {https://ui.adsabs.harvard.edu/abs/2010MNRAS.401.1029C} {401, 1029}

\bibitem[\protect\citeauthoryear{{Cumming}, {Butler}, {Marcy}, {Vogt}, {Wright}  \& {Fischer}}{{Cumming} et~al.}{2008}]{2008PASP..120..531C}
{Cumming} A.,  {Butler} R.~P.,  {Marcy} G.~W.,  {Vogt} S.~S.,  {Wright} J.~T.,   {Fischer} D.~A.,  2008, \mn@doi [\pasp] {10.1086/588487}, \href {https://ui.adsabs.harvard.edu/abs/2008PASP..120..531C} {120, 531}

\bibitem[\protect\citeauthoryear{{Demircan} \& {Kahraman}}{{Demircan} \& {Kahraman}}{1991}]{1991Ap&SS.181..313D}
{Demircan} O.,  {Kahraman} G.,  1991, \mn@doi [\apss] {10.1007/BF00639097}, \href {https://ui.adsabs.harvard.edu/abs/1991Ap&SS.181..313D} {181, 313}

\bibitem[\protect\citeauthoryear{{Dong} \& {Zhu}}{{Dong} \& {Zhu}}{2013}]{2013ApJ...778...53D}
{Dong} S.,  {Zhu} Z.,  2013, \mn@doi [\apj] {10.1088/0004-637X/778/1/53}, \href {https://ui.adsabs.harvard.edu/abs/2013ApJ...778...53D} {778, 53}

\bibitem[\protect\citeauthoryear{{Endl}, {K{\"u}rster}, {Els}, {Hatzes}, {Cochran}, {Dennerl}  \& {D{\"o}bereiner}}{{Endl} et~al.}{2002}]{2002A&A...392..671E}
{Endl} M.,  {K{\"u}rster} M.,  {Els} S.,  {Hatzes} A.~P.,  {Cochran} W.~D.,  {Dennerl} K.,   {D{\"o}bereiner} S.,  2002, \mn@doi [\aap] {10.1051/0004-6361:20020937}, \href {https://ui.adsabs.harvard.edu/abs/2002A&A...392..671E} {392, 671}

\bibitem[\protect\citeauthoryear{{Fernandes}, {Mulders}, {Pascucci}, {Mordasini}  \& {Emsenhuber}}{{Fernandes} et~al.}{2019}]{2019ApJ...874...81F}
{Fernandes} R.~B.,  {Mulders} G.~D.,  {Pascucci} I.,  {Mordasini} C.,   {Emsenhuber} A.,  2019, \mn@doi [\apj] {10.3847/1538-4357/ab0300}, \href {https://ui.adsabs.harvard.edu/abs/2019ApJ...874...81F} {874, 81}

\bibitem[\protect\citeauthoryear{{Fischer} \& {Valenti}}{{Fischer} \& {Valenti}}{2005}]{2005ApJ...622.1102F}
{Fischer} D.~A.,  {Valenti} J.,  2005, \mn@doi [\apj] {10.1086/428383}, \href {https://ui.adsabs.harvard.edu/abs/2005ApJ...622.1102F} {622, 1102}

\bibitem[\protect\citeauthoryear{{Ford} \& {Rasio}}{{Ford} \& {Rasio}}{2008}]{2008ApJ...686..621F}
{Ford} E.~B.,  {Rasio} F.~A.,  2008, \mn@doi [\apj] {10.1086/590926}, \href {https://ui.adsabs.harvard.edu/abs/2008ApJ...686..621F} {686, 621}

\bibitem[\protect\citeauthoryear{{Fulton} et~al.,}{{Fulton} et~al.}{2021}]{2021ApJS..255...14F}
{Fulton} B.~J.,  et~al., 2021, \mn@doi [\apjs] {10.3847/1538-4365/abfcc1}, \href {https://ui.adsabs.harvard.edu/abs/2021ApJS..255...14F} {255, 14}

\bibitem[\protect\citeauthoryear{{Gaia Collaboration}}{{Gaia Collaboration}}{2022}]{2022yCat.1355....0G}
{Gaia Collaboration} 2022, VizieR Online Data Catalog, \href {https://ui.adsabs.harvard.edu/abs/2022yCat.1355....0G} {p. I/355}

\bibitem[\protect\citeauthoryear{{Gaia Collaboration} et~al.,}{{Gaia Collaboration} et~al.}{2018}]{2018A&A...616A...1G}
{Gaia Collaboration} et~al., 2018, \mn@doi [\aap] {10.1051/0004-6361/201833051}, \href {https://ui.adsabs.harvard.edu/abs/2018A&A...616A...1G} {616, A1}

\bibitem[\protect\citeauthoryear{{Ghezzi}, {Montet}  \& {Johnson}}{{Ghezzi} et~al.}{2018}]{2018ApJ...860..109G}
{Ghezzi} L.,  {Montet} B.~T.,   {Johnson} J.~A.,  2018, \mn@doi [\apj] {10.3847/1538-4357/aac37c}, \href {https://ui.adsabs.harvard.edu/abs/2018ApJ...860..109G} {860, 109}

\bibitem[\protect\citeauthoryear{{Goldreich} \& {Tremaine}}{{Goldreich} \& {Tremaine}}{1980}]{1980ApJ...241..425G}
{Goldreich} P.,  {Tremaine} S.,  1980, \mn@doi [\apj] {10.1086/158356}, \href {https://ui.adsabs.harvard.edu/abs/1980ApJ...241..425G} {241, 425}

\bibitem[\protect\citeauthoryear{{Hallatt} \& {Lee}}{{Hallatt} \& {Lee}}{2020}]{2020ApJ...904..134H}
{Hallatt} T.,  {Lee} E.~J.,  2020, \mn@doi [\apj] {10.3847/1538-4357/abc1d7}, \href {https://ui.adsabs.harvard.edu/abs/2020ApJ...904..134H} {904, 134}

\bibitem[\protect\citeauthoryear{{Hayashi}}{{Hayashi}}{1981}]{1981PThPS..70...35H}
{Hayashi} C.,  1981, \mn@doi [Progress of Theoretical Physics Supplement] {10.1143/PTPS.70.35}, \href {https://ui.adsabs.harvard.edu/abs/1981PThPS..70...35H} {70, 35}

\bibitem[\protect\citeauthoryear{{Howard} et~al.,}{{Howard} et~al.}{2010}]{2010Sci...330..653H}
{Howard} A.~W.,  et~al., 2010, \mn@doi [Science] {10.1126/science.1194854}, \href {https://ui.adsabs.harvard.edu/abs/2010Sci...330..653H} {330, 653}

\bibitem[\protect\citeauthoryear{{Howard} et~al.,}{{Howard} et~al.}{2012}]{2012ApJS..201...15H}
{Howard} A.~W.,  et~al., 2012, \mn@doi [\apjs] {10.1088/0067-0049/201/2/15}, \href {https://ui.adsabs.harvard.edu/abs/2012ApJS..201...15H} {201, 15}

\bibitem[\protect\citeauthoryear{{Huang}, {Wu}  \& {Triaud}}{{Huang} et~al.}{2016}]{2016ApJ...825...98H}
{Huang} C.,  {Wu} Y.,   {Triaud} A. H.~M.~J.,  2016, \mn@doi [\apj] {10.3847/0004-637X/825/2/98}, \href {https://ui.adsabs.harvard.edu/abs/2016ApJ...825...98H} {825, 98}

\bibitem[\protect\citeauthoryear{{Ida} \& {Lin}}{{Ida} \& {Lin}}{2004}]{2004ApJ...616..567I}
{Ida} S.,  {Lin} D.~N.~C.,  2004, \mn@doi [\apj] {10.1086/424830}, \href {https://ui.adsabs.harvard.edu/abs/2004ApJ...616..567I} {616, 567}

\bibitem[\protect\citeauthoryear{{Johnson}, {Howard}, {Bowler}, {Henry}, {Marcy}, {Wright}, {Fischer}  \& {Isaacson}}{{Johnson} et~al.}{2010}]{2010PASP..122..701J}
{Johnson} J.~A.,  {Howard} A.~W.,  {Bowler} B.~P.,  {Henry} G.~W.,  {Marcy} G.~W.,  {Wright} J.~T.,  {Fischer} D.~A.,   {Isaacson} H.,  2010, \mn@doi [\pasp] {10.1086/653809}, \href {https://ui.adsabs.harvard.edu/abs/2010PASP..122..701J} {122, 701}

\bibitem[\protect\citeauthoryear{{Knutson} et~al.,}{{Knutson} et~al.}{2014}]{2014ApJ...785..126K}
{Knutson} H.~A.,  et~al., 2014, \mn@doi [\apj] {10.1088/0004-637X/785/2/126}, \href {https://ui.adsabs.harvard.edu/abs/2014ApJ...785..126K} {785, 126}

\bibitem[\protect\citeauthoryear{{Kozai}}{{Kozai}}{1962}]{1962AJ.....67..591K}
{Kozai} Y.,  1962, \mn@doi [\aj] {10.1086/108790}, \href {https://ui.adsabs.harvard.edu/abs/1962AJ.....67..591K} {67, 591}

\bibitem[\protect\citeauthoryear{{Kutra} \& {Wu}}{{Kutra} \& {Wu}}{2020}]{2020arXiv200308431K}
{Kutra} T.,  {Wu} Y.,  2020, arXiv e-prints, \href {https://ui.adsabs.harvard.edu/abs/2020arXiv200308431K} {p. arXiv:2003.08431}

\bibitem[\protect\citeauthoryear{{Lepot}, {Aubin}  \& {Clemens}}{{Lepot} et~al.}{2017}]{w9100796}
{Lepot} M.,  {Aubin} J.-B.,   {Clemens} F.~H.,  2017, \mn@doi [Interpolation in Time Series: An Introductive Overview of Existing Methods, Their Performance Criteria and Uncertainty Assessment] {10.3390/w9100796}, \href {https://www.mdpi.com/2073-4441/9/10/796} {9, 796}

\bibitem[\protect\citeauthoryear{{Lidov}}{{Lidov}}{1962}]{1962P&SS....9..719L}
{Lidov} M.~L.,  1962, \mn@doi [\planss] {10.1016/0032-0633(62)90129-0}, \href {https://ui.adsabs.harvard.edu/abs/1962P&SS....9..719L} {9, 719}

\bibitem[\protect\citeauthoryear{{Lin} \& {Papaloizou}}{{Lin} \& {Papaloizou}}{1986}]{1986ApJ...309..846L}
{Lin} D.~N.~C.,  {Papaloizou} J.,  1986, \mn@doi [\apj] {10.1086/164653}, \href {https://ui.adsabs.harvard.edu/abs/1986ApJ...309..846L} {309, 846}

\bibitem[\protect\citeauthoryear{{Lin}, {Bodenheimer}  \& {Richardson}}{{Lin} et~al.}{1996}]{1996Natur.380..606L}
{Lin} D.~N.~C.,  {Bodenheimer} P.,   {Richardson} D.~C.,  1996, \mn@doi [\nat] {10.1038/380606a0}, \href {https://ui.adsabs.harvard.edu/abs/1996Natur.380..606L} {380, 606}

\bibitem[\protect\citeauthoryear{{Lu}, {Angus}, {Curtis}, {David}  \& {Kiman}}{{Lu} et~al.}{2021}]{2021AJ....161..189L}
{Lu} Y.~L.,  {Angus} R.,  {Curtis} J.~L.,  {David} T.~J.,   {Kiman} R.,  2021, \mn@doi [\aj] {10.3847/1538-3881/abe4d6}, \href {https://ui.adsabs.harvard.edu/abs/2021AJ....161..189L} {161, 189}

\bibitem[\protect\citeauthoryear{{Mayor} et~al.,}{{Mayor} et~al.}{2011}]{2011arXiv1109.2497M}
{Mayor} M.,  et~al., 2011, arXiv e-prints, \href {https://ui.adsabs.harvard.edu/abs/2011arXiv1109.2497M} {p. arXiv:1109.2497}

\bibitem[\protect\citeauthoryear{{Mortier}, {Santos}, {Sousa}, {Israelian}, {Mayor}  \& {Udry}}{{Mortier} et~al.}{2013}]{2013A&A...551A.112M}
{Mortier} A.,  {Santos} N.~C.,  {Sousa} S.,  {Israelian} G.,  {Mayor} M.,   {Udry} S.,  2013, \mn@doi [\aap] {10.1051/0004-6361/201220707}, \href {https://ui.adsabs.harvard.edu/abs/2013A&A...551A.112M} {551, A112}

\bibitem[\protect\citeauthoryear{{Mulders}, {Pascucci}  \& {Apai}}{{Mulders} et~al.}{2015}]{2015ApJ...798..112M}
{Mulders} G.~D.,  {Pascucci} I.,   {Apai} D.,  2015, \mn@doi [\apj] {10.1088/0004-637X/798/2/112}, \href {https://ui.adsabs.harvard.edu/abs/2015ApJ...798..112M} {798, 112}

\bibitem[\protect\citeauthoryear{{Mulders}, {Pascucci}, {Apai}  \& {Ciesla}}{{Mulders} et~al.}{2018}]{2018AJ....156...24M}
{Mulders} G.~D.,  {Pascucci} I.,  {Apai} D.,   {Ciesla} F.~J.,  2018, \mn@doi [\aj] {10.3847/1538-3881/aac5ea}, \href {https://ui.adsabs.harvard.edu/abs/2018AJ....156...24M} {156, 24}

\bibitem[\protect\citeauthoryear{{Mulders}, {Pascucci}, {Ciesla}  \& {Fernandes}}{{Mulders} et~al.}{2021}]{2021AAS...23731702M}
{Mulders} G.~D.,  {Pascucci} I.,  {Ciesla} F.~J.,   {Fernandes} R.~B.,  2021, in American Astronomical Society Meeting Abstracts. p. 317.02

\bibitem[\protect\citeauthoryear{{Narang}, {Manoj}, {Furlan}, {Mordasini}, {Henning}, {Mathew}, {Banyal}  \& {Sivarani}}{{Narang} et~al.}{2018}]{2018AJ....156..221N}
{Narang} M.,  {Manoj} P.,  {Furlan} E.,  {Mordasini} C.,  {Henning} T.,  {Mathew} B.,  {Banyal} R.~K.,   {Sivarani} T.,  2018, \mn@doi [\aj] {10.3847/1538-3881/aae391}, \href {https://ui.adsabs.harvard.edu/abs/2018AJ....156..221N} {156, 221}

\bibitem[\protect\citeauthoryear{{Neves}, {Bonfils}, {Santos}, {Delfosse}, {Forveille}, {Allard}  \& {Udry}}{{Neves} et~al.}{2013}]{2013A&A...551A..36N}
{Neves} V.,  {Bonfils} X.,  {Santos} N.~C.,  {Delfosse} X.,  {Forveille} T.,  {Allard} F.,   {Udry} S.,  2013, \mn@doi [\aap] {10.1051/0004-6361/201220574}, \href {https://ui.adsabs.harvard.edu/abs/2013A&A...551A..36N} {551, A36}

\bibitem[\protect\citeauthoryear{{Ngo} et~al.,}{{Ngo} et~al.}{2015}]{2015ApJ...800..138N}
{Ngo} H.,  et~al., 2015, \mn@doi [\apj] {10.1088/0004-637X/800/2/138}, \href {https://ui.adsabs.harvard.edu/abs/2015ApJ...800..138N} {800, 138}

\bibitem[\protect\citeauthoryear{{Pascucci} et~al.,}{{Pascucci} et~al.}{2016}]{2016ApJ...831..125P}
{Pascucci} I.,  et~al., 2016, \mn@doi [\apj] {10.3847/0004-637X/831/2/125}, \href {https://ui.adsabs.harvard.edu/abs/2016ApJ...831..125P} {831, 125}

\bibitem[\protect\citeauthoryear{{Petigura} et~al.,}{{Petigura} et~al.}{2017}]{2017AJ....154..107P}
{Petigura} E.~A.,  et~al., 2017, \mn@doi [\aj] {10.3847/1538-3881/aa80de}, \href {https://ui.adsabs.harvard.edu/abs/2017AJ....154..107P} {154, 107}

\bibitem[\protect\citeauthoryear{{Petigura} et~al.,}{{Petigura} et~al.}{2018}]{2018AJ....155...89P}
{Petigura} E.~A.,  et~al., 2018, \mn@doi [\aj] {10.3847/1538-3881/aaa54c}, \href {https://ui.adsabs.harvard.edu/abs/2018AJ....155...89P} {155, 89}

\bibitem[\protect\citeauthoryear{{Pollack}, {Hubickyj}, {Bodenheimer}, {Lissauer}, {Podolak}  \& {Greenzweig}}{{Pollack} et~al.}{1996}]{1996Icar..124...62P}
{Pollack} J.~B.,  {Hubickyj} O.,  {Bodenheimer} P.,  {Lissauer} J.~J.,  {Podolak} M.,   {Greenzweig} Y.,  1996, \mn@doi [\icarus] {10.1006/icar.1996.0190}, \href {https://ui.adsabs.harvard.edu/abs/1996Icar..124...62P} {124, 62}

\bibitem[\protect\citeauthoryear{{Rasio} \& {Ford}}{{Rasio} \& {Ford}}{1996}]{1996Sci...274..954R}
{Rasio} F.~A.,  {Ford} E.~B.,  1996, \mn@doi [Science] {10.1126/science.274.5289.954}, \href {https://ui.adsabs.harvard.edu/abs/1996Sci...274..954R} {274, 954}

\bibitem[\protect\citeauthoryear{{Rosenthal} et~al.,}{{Rosenthal} et~al.}{2021}]{2021ApJS..255....8R}
{Rosenthal} L.~J.,  et~al., 2021, \mn@doi [\apjs] {10.3847/1538-4365/abe23c}, \href {https://ui.adsabs.harvard.edu/abs/2021ApJS..255....8R} {255, 8}

\bibitem[\protect\citeauthoryear{{Santerne} et~al.,}{{Santerne} et~al.}{2016}]{2016A&A...587A..64S}
{Santerne} A.,  et~al., 2016, \mn@doi [\aap] {10.1051/0004-6361/201527329}, \href {https://ui.adsabs.harvard.edu/abs/2016A&A...587A..64S} {587, A64}

\bibitem[\protect\citeauthoryear{{Steffen} et~al.,}{{Steffen} et~al.}{2012}]{2012PNAS..109.7982S}
{Steffen} J.~H.,  et~al., 2012, \mn@doi [Proceedings of the National Academy of Science] {10.1073/pnas.1120970109}, \href {https://ui.adsabs.harvard.edu/abs/2012PNAS..109.7982S} {109, 7982}

\bibitem[\protect\citeauthoryear{{Su}, {Xie}, {Zhou}  \& {Thebault}}{{Su} et~al.}{2021}]{2021AJ....162..272S}
{Su} X.-N.,  {Xie} J.-W.,  {Zhou} J.-L.,   {Thebault} P.,  2021, \mn@doi [\aj] {10.3847/1538-3881/ac2ba3}, \href {https://ui.adsabs.harvard.edu/abs/2021AJ....162..272S} {162, 272}

\bibitem[\protect\citeauthoryear{{Thompson} et~al.,}{{Thompson} et~al.}{2018}]{2018ApJS..235...38T}
{Thompson} S.~E.,  et~al., 2018, \mn@doi [\apjs] {10.3847/1538-4365/aab4f9}, \href {https://ui.adsabs.harvard.edu/abs/2018ApJS..235...38T} {235, 38}

\bibitem[\protect\citeauthoryear{{Weidenschilling} \& {Marzari}}{{Weidenschilling} \& {Marzari}}{1996}]{1996Natur.384..619W}
{Weidenschilling} S.~J.,  {Marzari} F.,  1996, \mn@doi [\nat] {10.1038/384619a0}, \href {https://ui.adsabs.harvard.edu/abs/1996Natur.384..619W} {384, 619}

\bibitem[\protect\citeauthoryear{{Wittenmyer} et~al.,}{{Wittenmyer} et~al.}{2016}]{2016ApJ...819...28W}
{Wittenmyer} R.~A.,  et~al., 2016, \mn@doi [\apj] {10.3847/0004-637X/819/1/28}, \href {https://ui.adsabs.harvard.edu/abs/2016ApJ...819...28W} {819, 28}

\bibitem[\protect\citeauthoryear{{Wittenmyer} et~al.,}{{Wittenmyer} et~al.}{2020a}]{2020MNRAS.491.5248W}
{Wittenmyer} R.~A.,  et~al., 2020a, \mn@doi [\mnras] {10.1093/mnras/stz3378}, \href {https://ui.adsabs.harvard.edu/abs/2020MNRAS.491.5248W} {491, 5248}

\bibitem[\protect\citeauthoryear{{Wittenmyer} et~al.,}{{Wittenmyer} et~al.}{2020b}]{2020MNRAS.492..377W}
{Wittenmyer} R.~A.,  et~al., 2020b, \mn@doi [\mnras] {10.1093/mnras/stz3436}, \href {https://ui.adsabs.harvard.edu/abs/2020MNRAS.492..377W} {492, 377}

\bibitem[\protect\citeauthoryear{{Yang}, {Xie}  \& {Zhou}}{{Yang} et~al.}{2020}]{2020AJ....159..164Y}
{Yang} J.-Y.,  {Xie} J.-W.,   {Zhou} J.-L.,  2020, \mn@doi [\aj] {10.3847/1538-3881/ab7373}, \href {https://ui.adsabs.harvard.edu/abs/2020AJ....159..164Y} {159, 164}

\makeatother
\end{thebibliography}



\appendix

\section{The relation of stellar mass and effective temperature }
\label{sec:a1}

\begin{figure*}
	\centering
	\includegraphics[width =0.9 \textwidth]{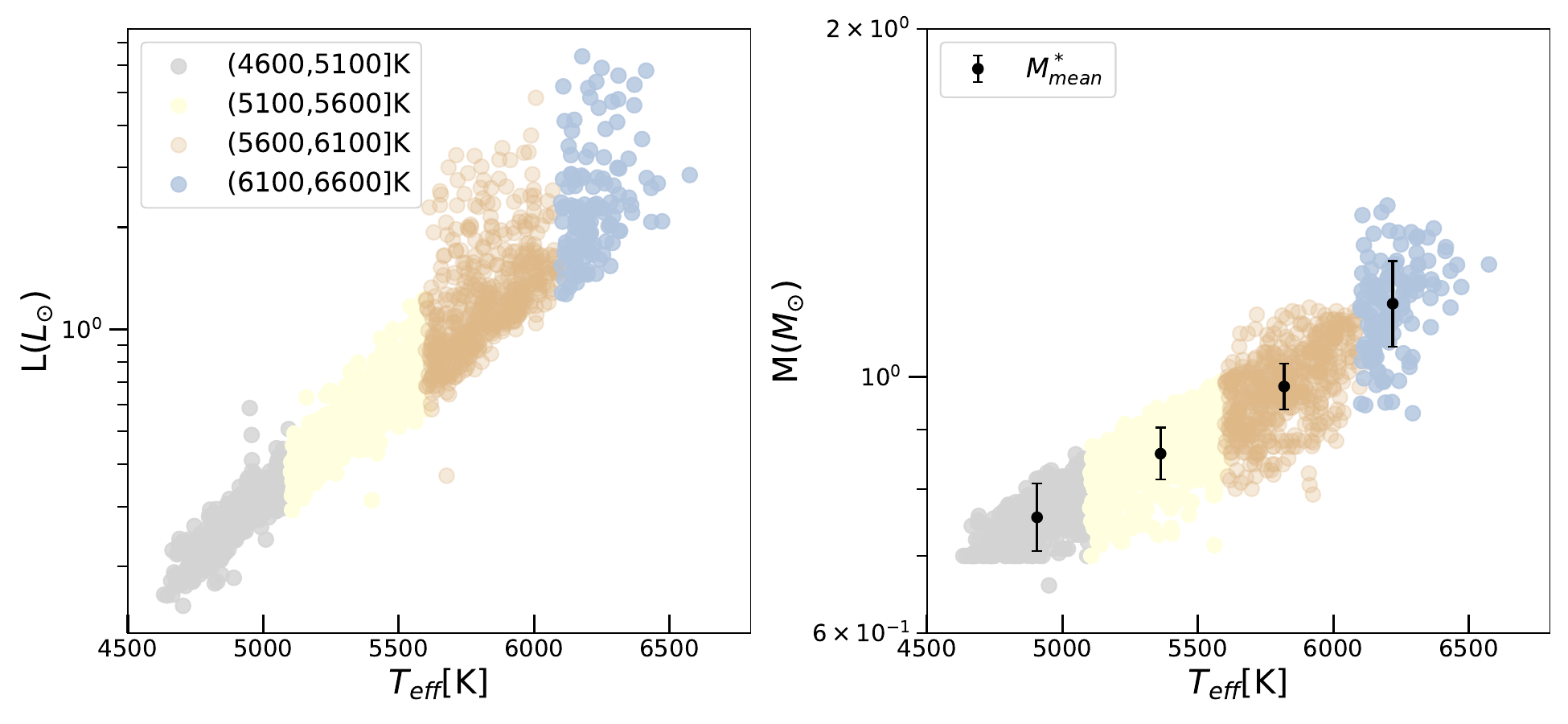}
	
	\caption{ $T_{\rm eff}$-mass diagrams for our star samples, different colors represent different bins based on effective temperature. Black dot represent the mean mass of each $T_{\rm eff}$ bin, the error bars indicate the $68.3\%$ confidence intervals.
	\label{fig:T_MS}}
\end{figure*}

For stars in the main sequence stage, there exists a general trend where a decrease in effective temperature corresponds to a decrease in the mass of the star. However, it is important to note that this relationship is not strictly one-to-one due to variations in the age and composition of the star. The left panel of Figure \ref{fig:T_MS} is the H-R diagram of our stellar samples, most stars are in the main sequence stage.
The right panel of Figure \ref{fig:T_MS} shows the mass and effective temperature for stars in our star sample. As is shown in the picture, for most stars, the higher the effective temperature, the more massive. And the mean mass of each $T_{\rm eff}$ bin increases monotonically with the increase of effective temperature.

\section{Comparison of effective temperature data}
\label{sec:ab}

\begin{figure*}
	\centering
	\includegraphics[width =0.9 \textwidth]{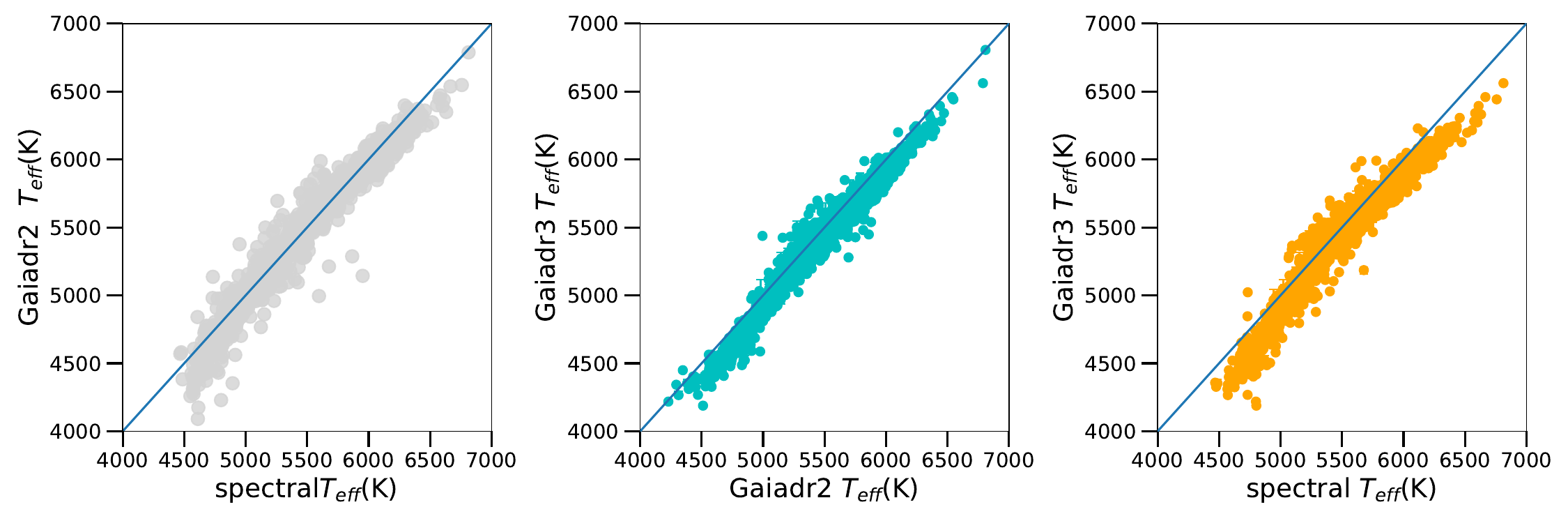}
	
	\caption{Comparison between effective temperatures of stars samples in GAIA DR3, GAIA DR2 and the raw stellar effective temperatures inferred through spectroscopic analysis.
	\label{fig:T_VS}}
\end{figure*}

To compare the effective temperature data, we examined multiple sources and analyzed their consistency. The results are summarized below. Specifically, we examine the effective temperature values derived from different data sets, including the raw stellar effective temperatures inferred through spectroscopic analysis\citep{2013A&A...551A.112M}, GAIA DR2\citep{2018A&A...616A...1G} and GAIA DR3\citep{2022yCat.1355....0G}. The effective temperature of stars in GAIA DR2 is estimated from Apsis-Priam, and the effective temperature in GAIA DR3 is inferred by GSP-Phot Aeneas from BP/RP spectra. The detailed comparison of the effective temperature data is presented in Figure \ref{fig:T_VS}. As shown in the first columns of Figure \ref{fig:T_VS}, The disparity between the effective temperature derived from GAIA DR2 and the effective temperature data inferred through spectroscopic analysis is comparatively small. The last two columns of Figure \ref{fig:T_VS} compare the effective temperature in GAIA DR3 to GAIA DR2 and raw data, respectively.
$T_{\rm eff}$ of stars in GAIA DR2 and raw data are warmer than the effective temperature in GAIA DR3. Our analysis reveals that the effective temperature values obtained from Gaia DR2 data are consistent within the margin of error with both Gaia DR3 and the spectroscopic analysis results. This indicates that the previously used Gaia DR2 data remains reliable for our purposes.

\section{Testing the Sensitivity of Occurrence Rates and Mutual Occurrence Ratios of Giant Planets to Boundary Adjustments}
\label{sec:ac}

\begin{figure*}
	\centering
	\includegraphics[width =0.9 \textwidth]{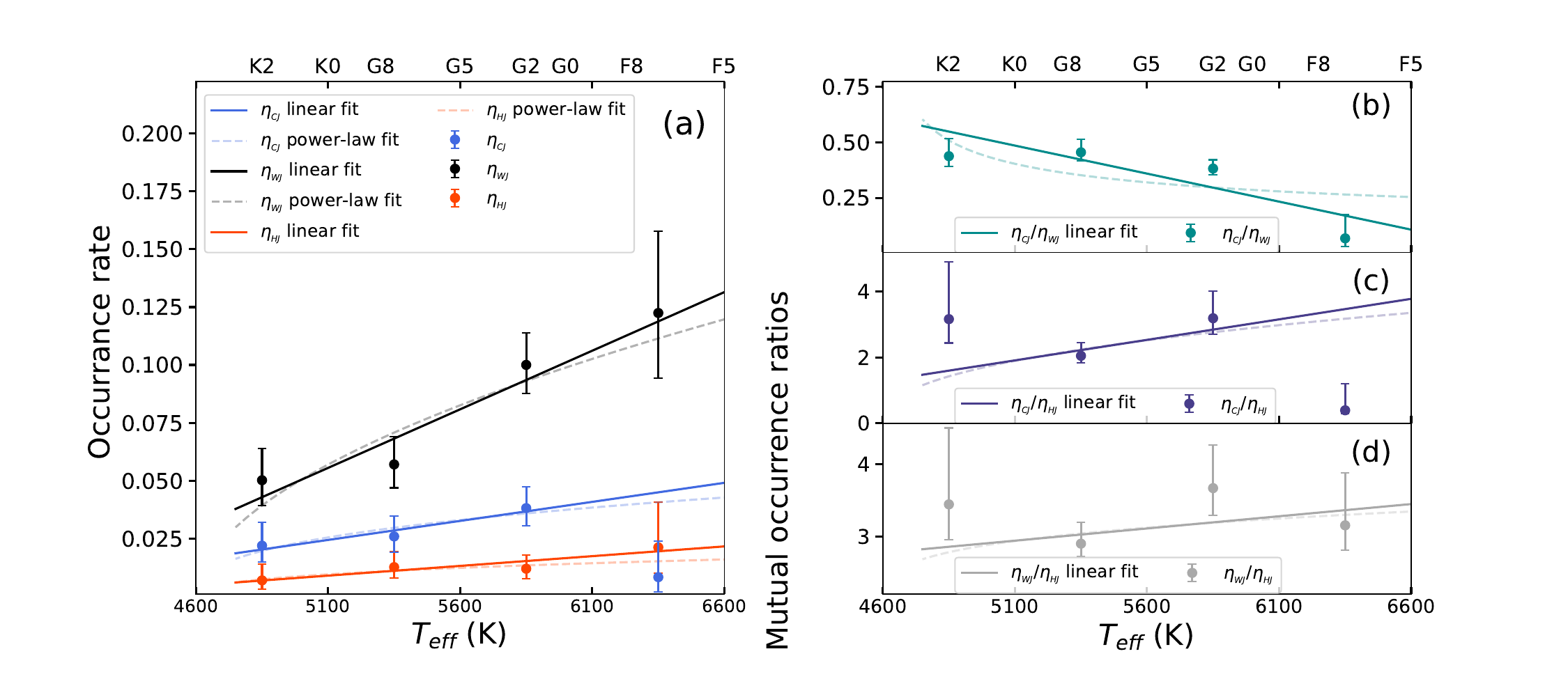}
	
	\caption{Similar to Figure \ref{fig:RV_all}, but here, giant planet with $0.08<{a_{\rm p}}/a_{\rm snow}<1.0$ as WJ .
	\label{fig:0.08}}
	\end{figure*}

\begin{figure*}
	\centering
	\includegraphics[width =0.9 \textwidth]{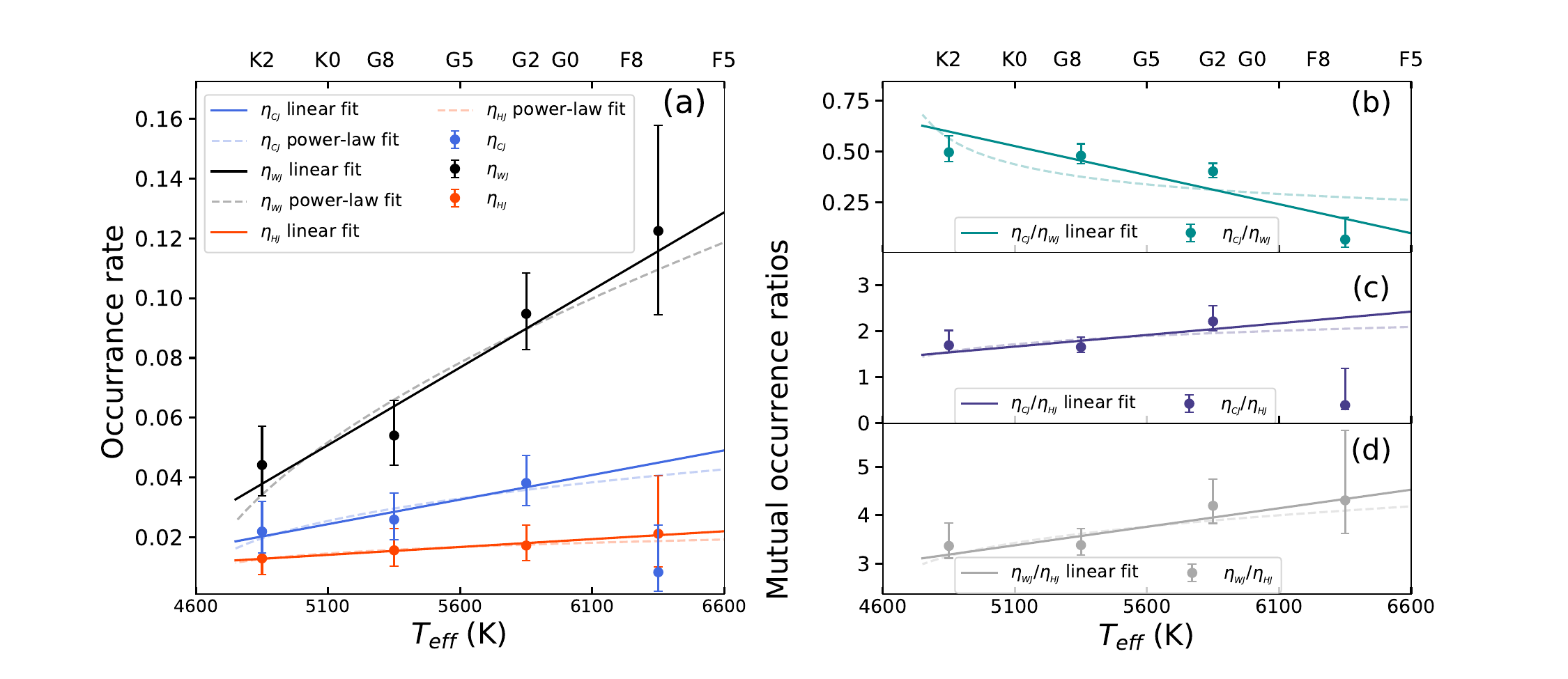}
	
	\caption{Similar to Figure \ref{fig:RV_all}, but here, giant planet with $0.12<{a_{\rm p}}/a_{\rm snow}<1.0$ as WJ.
	\label{fig:0.12}}\end{figure*}

To assess the sensitivity of occurrence rates and mutual occurrence ratios of giant planets to adjustments, we conducted two analyses using various threshold values, such as $0.08<{a_{\rm p}}/a_{\rm snow}<1.0$ and $0.12<{a_{\rm p}}/a_{\rm snow}<1.0$ as the boundary of WJ. 0.08 and 0.12 are taken around 0.1. As can be seen in panel (a) of Figure \ref{fig:0.08} and Figure \ref{fig:0.12}, with the increase of effective temperature, the occurrence rate all showed an increasing trend, Similar to the result  in section \ref{sec:results}, but the trend is different from the trend shown in Figure \ref{fig:RV_all}. The results demonstrate that the occurrence rates of giant planets are not significantly influenced by changes in the boundary criteria. For mutual occurrence ratios of HJ, WJ and CJ, the panel (b) of Figure \ref{fig:0.08} and \ref{fig:0.12} shows ${\eta}{_{\rm CJ}}/{\eta}_{\rm WJ}$ still shows a consistent downward trend with the increasing stellar effective temperature. As can be seen in panel (c) of  \ref{fig:0.12}, ${\eta}{_{\rm CJ}}/{\eta}_{\rm HJ}$ also shows the opposite trend from panel (b), it is similar to the result  in section \ref{sec:results}. And the trend of ${\eta}{_{\rm WJ}}/{\eta}_{\rm HJ}$ show in the panel (d) of Figure \ref{fig:0.12} also have a increasing trend with the increasing $T_{\rm eff}$. But the trend of ${\eta}{_{\rm CJ}}/{\eta}_{\rm HJ}$ and ${\eta}{_{\rm WJ}}/{\eta}_{\rm HJ}$ show in the panel (c) and (d) of Figure \ref{fig:0.08} is different to the panel (c) and (d) of Figure \ref{fig:RV_all}. Changes in the boundary criteria of WJ and HJ significantly influence the mutual occurrence ratios of HJ and WJ, but mutual occurrence ratios ${\eta}{_{\rm CJ}}/{\eta}_{\rm WJ}$ not sensitive to the definition of WJ.

\section{A boundary of HJ and WJ in Kepler DR25}
\label{sec:kepler}	
\begin{figure*}
	\centering
	\includegraphics[width =0.9 \textwidth]{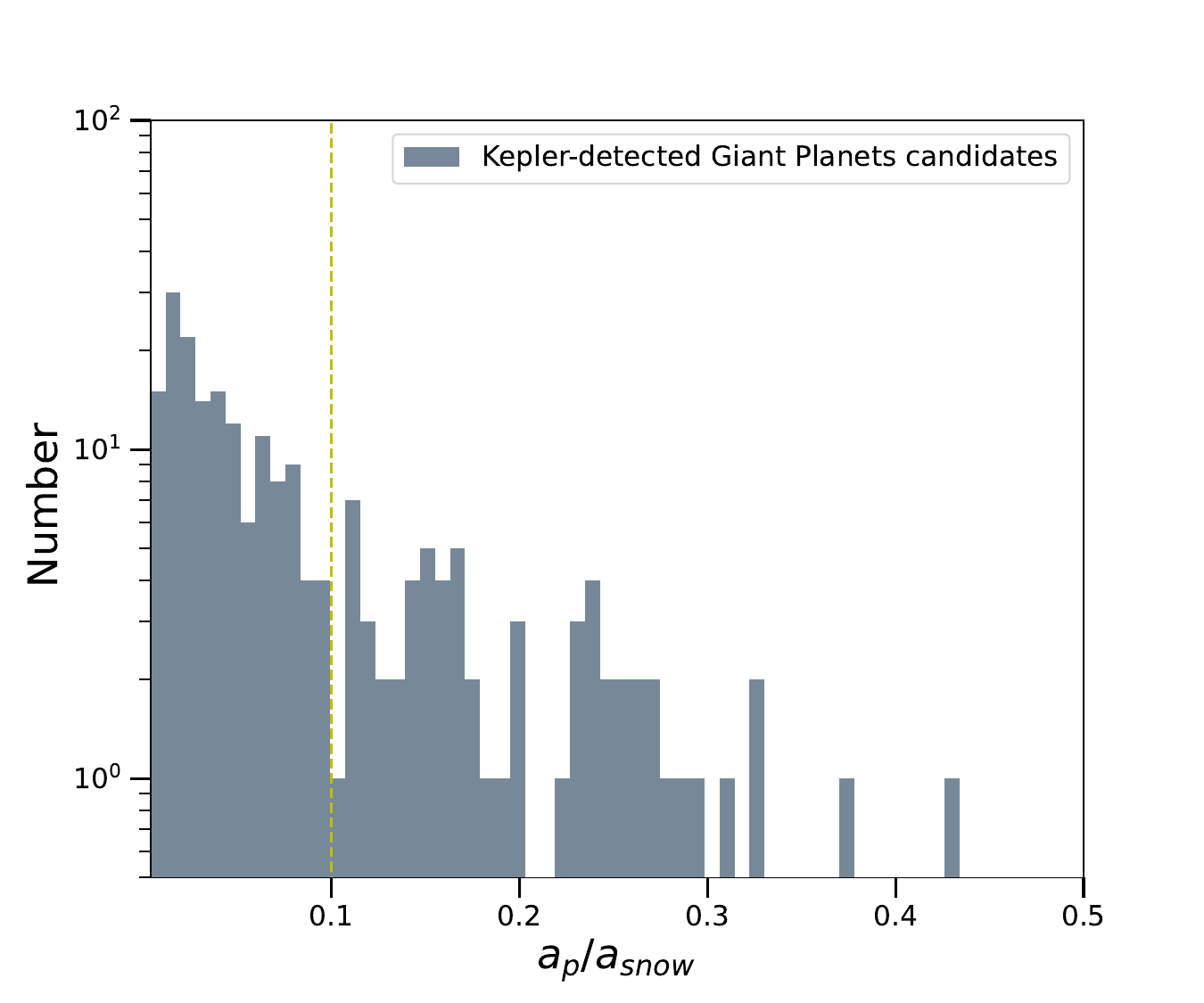}
	
	\caption{Similar to panel (b) of Figure \ref{fig:RV_b}, but here, giant planet samples data from Kepler DR25. There is also a valley around ${a_{\rm p}}/a_{\rm snow}\sim 0.1$.
	\label{fig:kepler}}\end{figure*}

The Kepler DR25 also offers a homogeneous sample of short- and moderate-period planets with survey completeness well determined \citep{2018AJ....156...24M, 2018ApJS..235...38T}. We adopt the stellar properties after across match with Gaia Data Release 2\citep{2020arXiv200514671B}, e.g. the effective temperature$T_{eff}$, stellar radius $R_{\star}$. With these well determined properties, and retain only main sequence stars . After updated Radius of planet according to previous mentioned stellar parameters(more detail see the second paper of this series). Only those giant planets with $R_{p}$ between $4R_{\oplus}$ and $20R_{\oplus}$ are selected. Calculating and counting the normalized orbit semi-major axis (${a_p}/a_{snow}$) of giant planet candidates, Shown in Figure \ref{fig:kepler}. As we can see in Figure \ref{fig:kepler}, a gap is exist in the bin around ${a_p}/a_{snow}=0.1$, it similar to the gap exist in RV sample. But there is a different that ${a_p}/a_{snow}=0.1$ located the right edge of RV sample but located the left edge of Kepler sample. So, we take ${a_p}/a_{snow}=0.1$ as the boundary of HJ and WJ. This is an approximate position, which corresponds to a planetary effective temperature of about 537K, planetary effective temperature is the physical quantity that more effectively integrates the description of planetary position and laminous flux desity.

{\Zhang \section{Analysis of the effect of planetary orbit eccentricity on results}
\label{sec:eccentricity}	
In our analysis of 124 giant planets, 122 exhibit orbital eccentricities greater than 0, with the majority displaying values below 0.5. For those with eccentricities beneath this threshold, \citet{2010MNRAS.401.1029C}
 have shown that velocity amplitude upper limits calculated for circular orbits can serve as a reliable proxy for the upper bounds of amplitude in eccentric orbits, provided the eccentricity (e) is approximately 0.5 or lower. Furthermore, \citet{2016ApJ...819...28W} found that approximations based on circular orbits are adequate for estimating the detection limits of Jupiter analogs in low-eccentricity orbits. Similarly,  \citet{2019ApJ...874...81F} applied the concept of detection completeness for circular orbits to assess the occurrence rates of giant planets. Nevertheless, our preliminary analysis explores the implications of orbital eccentricity on these findings.
 
High-eccentricity planets are relatively rare within our sample, however, some may traverse the snow line. To evaluate the influence of all non-zero eccentricities, we considered both the periastron $a_{per}=a_{p}(1-e)$ and apastron $a_{apa}=a_{p}(1+e)$ distances. We categorized Hot Jupiters (HJs) as having a normalized apastron distance $a_{apa} < 0.1$, Warm Jupiters (WJs) with $a_{apa} < 1$ and a periastron distance $a_{per} > 0.1$, and Cold Jupiters (CJs) with  $a_{per} >1$. Our findings indicate that as the effective stellar temperature increases, there is a noticeable variation in the occurrence rates and mutual occurrence ratios of HJs, WJs, and CJs, as depicted in Figure \ref{fig:e0}. Specifically, while the general trend in planetary occurrence rates mirrors that of Figure \ref{fig:RV_all}, the absolute values, especially for CJs, differ substantially. As demonstrated in Figure \ref{fig:e0}(c), mutual occurrence ratios of HJ and CJ (${\eta}{_{\rm CJ}}/{\eta}_{\rm HJ}$) rise with increasing stellar temperature, albeit to a lesser extent than previously documented. Conversely, Figure \ref{fig:e0}(b) and (d), reveal that the mutual occurrence ratios ${\eta}{_{\rm CJ}}/{\eta}_{\rm WJ}$ and ${\eta}{_{\rm WJ}}/{\eta}_{\rm HJ}$ do not exhibit a clear monotonic trend, contrasting with the patterns observed in Figure \ref{fig:RV_all}. These outcomes suggest that while the overall occurrence rate of giant planets may not be markedly influenced by orbital eccentricity, the presence of orbits intersecting the snow line significantly affects the computed mutual occurrence ratios.

\begin{figure*}
	\centering
	\includegraphics[width =0.9 \textwidth]{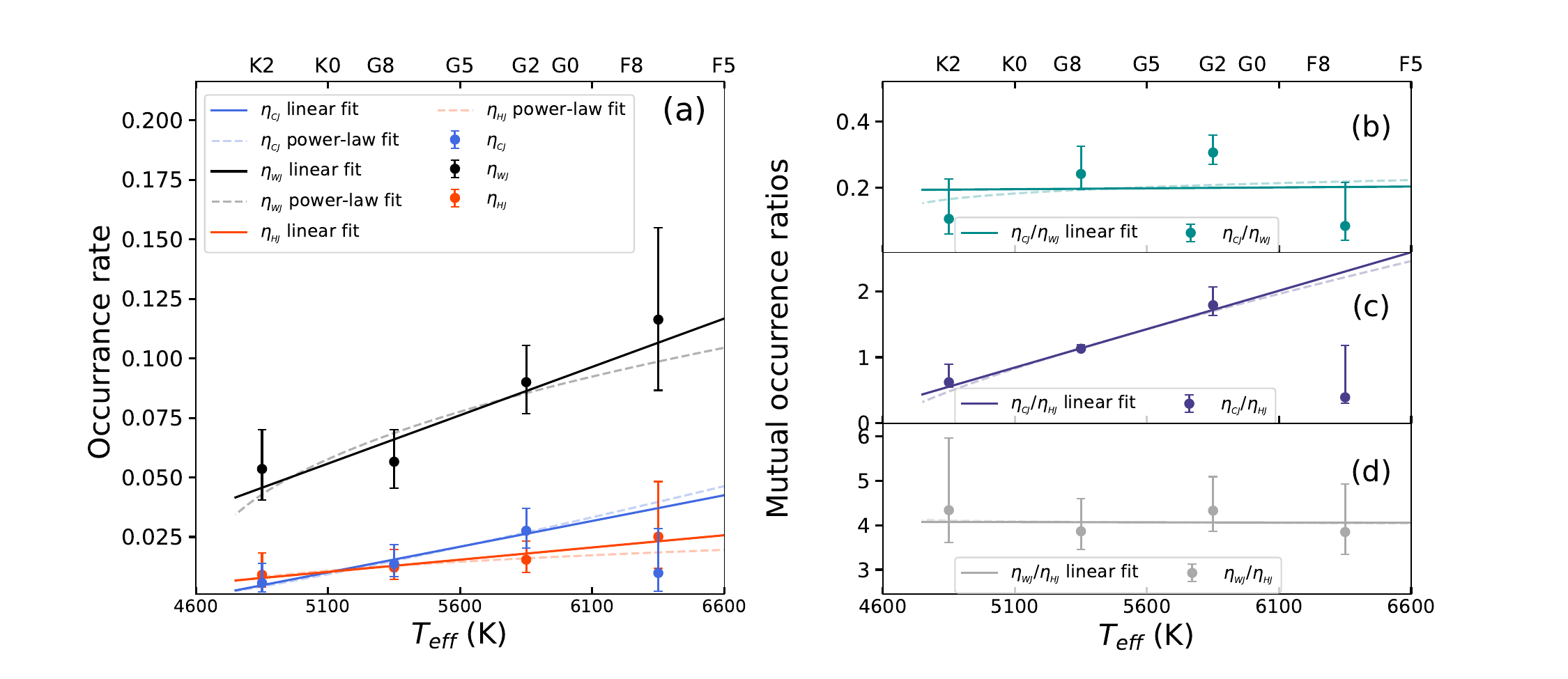}
	
	\caption{Similar to Figure \ref{fig:RV_all}, but here, giant planet with normalized apastron distance $a_{apa}<0.1$ as HJ,  WJ have $a_{per}>0.1$ and $a_{apa}<1$, and CJ have $a_{per}>1$.
	\label{fig:e0}}\end{figure*}
}

\bsp	
\label{lastpage}
\end{document}